\documentclass[times]{aastex631}
\usepackage{amsmath}
\usepackage{comment}
\usepackage{amsbsy}

\usepackage{booktabs}

\shorttitle{Toward Improved Understanding of Magnetic Fields Participating in Solar Flares}
\shortauthors{Kazachenko et al.}

\begin{document}
\title{Toward Improved Understanding of Magnetic Fields Participating in Solar Flares:  Statistical Analysis of Magnetic Field within Flare Ribbons}

\author[0000-0001-8975-7605]{Maria~D.~Kazachenko}
\affiliation{Dept. of Astrophysical and Planetary Sciences, University of Colorado Boulder,  \\
 2000 Colorado Ave, Boulder, CO 80305, USA}
\affiliation{National Solar Observatory,\\
 3665 Discovery Drive, Boulder, CO 80303, USA}

\author[0000-0001-6886-855X ]{Benjamin J. Lynch}
\affiliation{Space Sciences Laboratory, University of California, Berkeley, \\
CA 94720, USA}

\author[0000-0002-5598-046X]{Antonia Savcheva}
\affiliation{Harvard-Smithsonian Center for Astrophysics,\\
60 Garden st., Cambridge, MA 02138, USA}
\affiliation{Institute of Astronomy and National Astronomical Observatory, Bulgarian Academy of Sciences, \\
72 Tsarigradsko Chaussee Blvd., 1784 Sofia, Bulgaria}

\author[0000-0003-4043-616X ]{Xudong Sun}
\affiliation{Institute for Astronomy, University of Hawaii at Manoa, \\
34 Ohia Ku St, Pukalani, HI 96768, USA}

\author[0000-0003-2244-641X ]{Brian T. Welsch}
\affiliation{Dept. of Natural \& Applied Sciences, Univ. of Wisconsin - Green Bay, \\
Green Bay, WI 54311, USA}

\email{maria.kazachenko@colorado.edu}
\begin{abstract}
Violent solar flares and coronal mass ejections (CMEs) are magnetic phenomena. However, how magnetic fields reconnecting in the flare differ from non-flaring magnetic fields remains unclear owing to the lack of studies of the flare magnetic properties. Here we present a first statistical study of flaring (highlighted by flare-ribbons) vector magnetic fields in the photosphere. Our systematic approach allows us to describe key physical properties of solar flare magnetism, including distributions of magnetic flux, magnetic shear, vertical current and net current over flaring versus non-flaring parts of the active region, and compare these with flare/CME properties. Our analysis suggests that while flares are guided by the physical properties that scale with AR size, like the total amount of magnetic flux that participates in the reconnection process and the total current (extensive properties), CMEs are guided by mean properties, like the fraction of the AR magnetic flux that participates (intensive property), with little dependence on the amount of shear at polarity inversion line (PIL) or the net current. We find that the non-neutralized current is proportional to the amount of shear at PIL, providing direct evidence that net vertical currents are formed as a result of any mechanism that could generate magnetic shear along PIL. We also find that eruptive events tend to have smaller PIL fluxes and larger magnetic shears than confined events. Our analysis provides a reference for more realistic solar and stellar flare models. The database is available online and can be used for future quantitative studies of flare magnetism.

\end{abstract}
%
%

\section{Introduction}\label{intro}
\long\def\/*#1*/{}
%
Solar flares and coronal mass ejections (CMEs) that produce the most severe space weather disturbances arise from significant reconfiguration of magnetic fields in the solar corona.  Because these coronal fields are rooted in the underlying, much-denser solar photosphere, understanding how flares and CMEs work requires understanding the structure and evolution of magnetic fields from the photosphere to the corona. Historically, most of the analyses of photospheric magnetic fields before and during the flares focused on the active region (AR) as a whole (e.g. \citealt{Toriumi2019}). However, flare observations have shown that only a fraction of the AR magnetic field participates in the flare --  from the analysis of $3000$ solar flares we found that the fraction of active-region magnetic flux that undergoes reconnection ranges from $(3\pm2)\%$ to $(21\pm10)\%$ for C1 to X-class flares, respectively  (see Fig. 11 in \citealt{Kazachenko2017}). Moreover, subsequent flares can occur in very different areas of the same AR -- a well-known example is the two extreme events on 2003 October 28 and 29 that occurred in different locations in AR 10486 \citep{Kazachenko2010}. This implies that to understand why and how flares occur, we need to understand the properties of the magnetic fields that participate in the flare and also how they differ from the AR as a whole.

\begin{figure*}[htb!]
  \centering  
\resizebox{0.9\hsize}{!}{\includegraphics[angle=0,width=\textwidth]{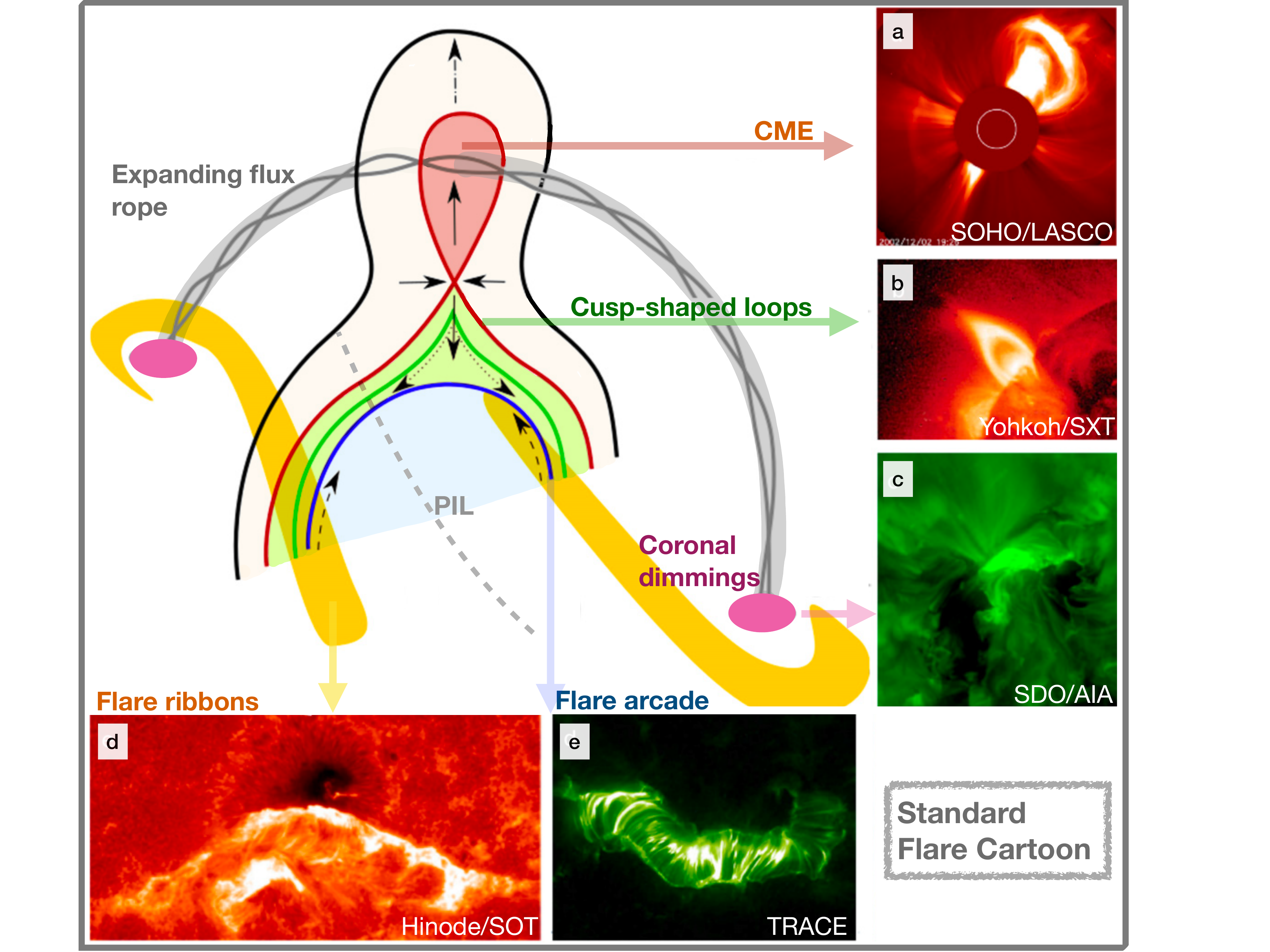}}  \caption{Standard flare cartoon from \citet{Savcheva2016}. As the flux rope rises (gray), it stretches the current sheet and the QSLs  in the plane perpendicular to the flux rope (red) forming a set of newly reconnected cusped and arcade-shaped field lines (green and blue, respectively) with footpoints, highlighted as flare ribbons (yellow). Images on the right and bottom show the main observed features explained by the model: (a) CME with three-part structure from SOHO/LASCO; (b) cusped loops from Yohkoh/SXT; (c) coronal dimmings from SDO/AIA; (d) flare ribbons from Hinode/SOT; and (e) flare arcade from TRACE. Published with the author's permission.}
  \label{cshkp}
\end{figure*}

According to the canonical, two-dimensional (2D) flare model called the CSHKP model \citep{Carmichael1964,Sturrock1968,Hirayama1974,Kopp1976} and its extension to 3D  \citep{Longcope2007,Aulanier2012,Janvier2014,Savcheva2015,Savcheva2016}, maps of flare ribbons identify footpoints of newly-reconnected magnetic fields (see Figure~\ref{cshkp}).
In relation to magnetic topology, ribbons correspond to locations where separatrices, dividing domains of distinct connectivity, like spines and fans \citep{Longcope2007,Kazachenko2012} or quasi-separatrix layers (QSL, 
\citealt{Savcheva2015,Savcheva2016}) intersect with the chromosphere. Until recently, limited accuracy of both vector-magnetic-field and ribbon measurements  made systematic analysis of magnetic fields within flare ribbons areas limited.

The launch of the {\it Solar Dynamics Observatory} \citep[SDO;][]{Pesnell2012}, with the
Helioseismic and Magnetic Imager \citep[HMI;][]{Scherrer2012,Hoeksema2014} and the Atmospheric Imaging Assembly \citep[AIA;][]{Lemen2012} instruments, marks the first time when both a vector magnetograph and ribbon-imaging capabilities became available on the same observing platform, making co-registration of AIA and HMI full-disk data relatively easy. In this paper, we make use of these two instruments and present a database of vector magnetic field properties swept by flare ribbons in $40$ flaring events, C-class and larger. Our intentions are twofold. First, we provide the reference information for the data set by describing the key processing procedures. Second, we present statistical analyses of the photospheric magnetic field participating in the flare and their relationship with that of the polarity inversion line (PIL) and the AR as a whole.

This paper follows the study of \citet{Kazachenko2017} that presented a \verb+RibbonDB+ database of $3137$ solar flare ribbon events, their   ribbon area and ribbon magnetic fluxes (reconnection fluxes). Complementing our previous study, here we create a new dataset, \verb+FlareMagDB+, and describe the vector magnetic fields properties: magnetic flux, mean magnetic  shear, vertical current and the net current within flare ribbons, AR and PIL areas.

This paper is organized as follows. In \S\ref{data}, we describe the SDO data we use, the list of selected events, and the methods. In \S\ref{res01}, we describe statistical results of the \verb+FlareMagDB+ analysis. In \S\ref{prelim-mhd}, we describe the results of the analysis of the 3D MHD ARMS simulation. Finally, in \S\ref{disc} and \S\ref{conc}, we discuss the results and summarize conclusions.

\section{Data and Methods}\label{data}
In this section, we describe the data and the methods we used to identify various physical properties of the AR magnetic field: identification of the polarity inversion line (PILs), flare ribbons and AR areas;  estimates of the magnetic fluxes, shears, vertical currents and net currents within PIL, ribbons and AR areas.
\subsection{Data}
\begin{figure*}[htb!]
  \centering  
\resizebox{1.0\hsize}{!}{\includegraphics[angle=0,width=\textwidth]{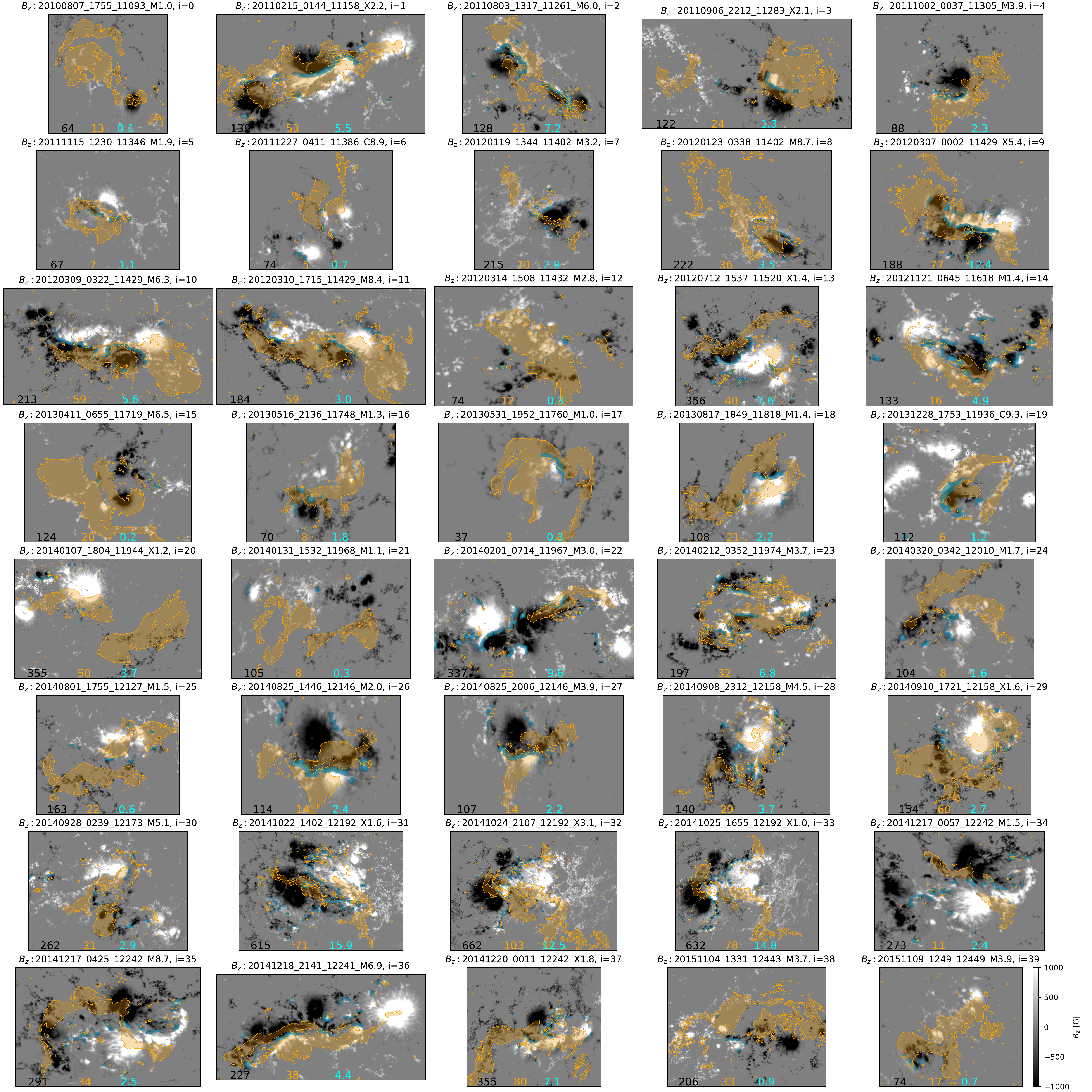}}  \caption{Vertical magnetic field maps, $B_z$, for $40$ events from the FlareMagDB database. Cyan color shows PIL areas. Orange color shows cumulative ribbon areas. Labels above each figure combine event's index in the database, flare start time, NOAA active region number and the GOES peak X-ray flux. ``Eruptive/confined'' label indicates whether the event is eruptive or confined. Black, orange and cyan numbers indicate total magnetic flux within AR, ribbon and PIL areas, respectively, in units of $10^{20}$ Mx (see column $\Phi$ in Table~\ref{events}). Here and in Figures~\ref{shear_examples} and \ref{jz_examples}, we zoom into the ROI's field-of-view to highlight the AR structure.}
  \label{bz_examples}
\end{figure*}

\begin{figure*}[tb!]
  \centering
\resizebox{1.0\hsize}{!}{\includegraphics[angle=0]{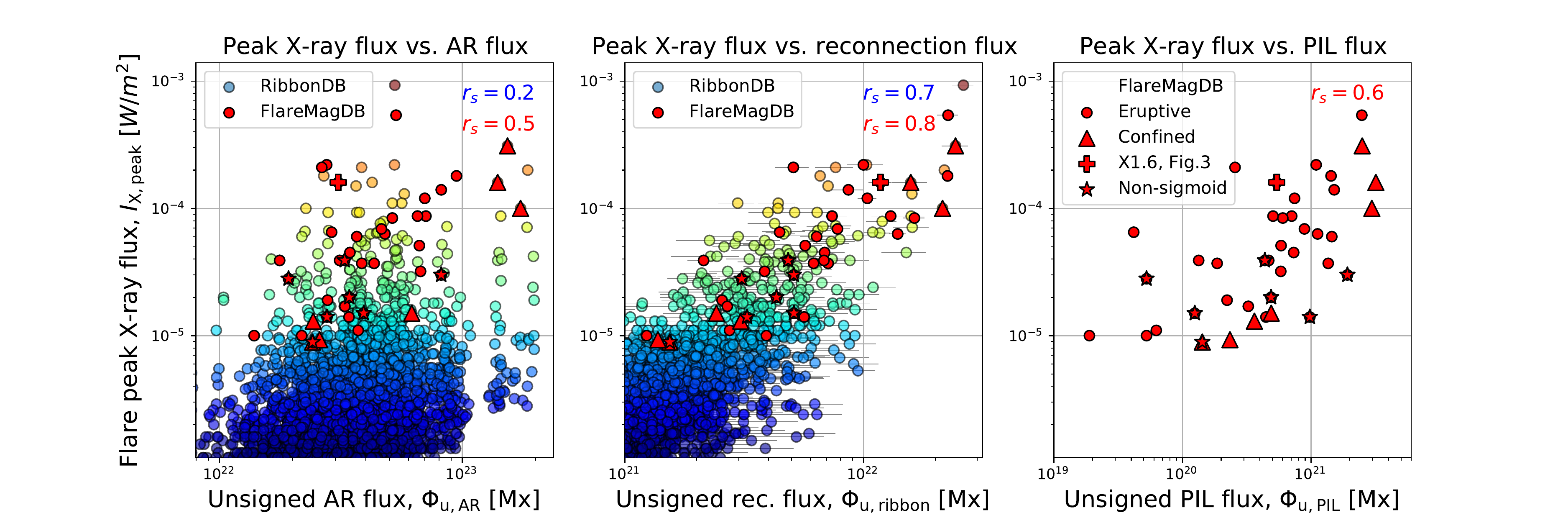}}
  \caption{Comparison of FlareMagDB (red) vs. RibbonDB (rainbow) flare samples: scatter plots of peak X-ray flux vs. unsigned pre-flare AR magnetic flux ({\it left}),  flare reconnection flux  ({\it middle}) and flux within the PIL ({\it right}). {\it Rainbow color} shows peak X-ray fluxes of events from the RibbonDB database; it varies from blue  for C-class flares to orange for X-class flares. {\it Red color shows} a subset of $40$ FlareMagDB  events. A ``+''-sign indicates the example X1.6 event in AR 12158 (see Figure~\ref{example}); $\bigtriangleup$ show $7$ non-CME events.  Stars show events that do not exhibit sigmoidal structure in EUV images. See \S\ref{data}.}
  \label{db}
\end{figure*}

We have assembled a  \verb+FlareMagDB+ catalogue that includes $40$ flares: $10$ X-, $28$ M- and $2$ C-class flares (see Table~\ref{events}). To give an overview of these $40$ events, in Figure~\ref{bz_examples} we show vertical magnetic field maps of $40$ ARs hosting these flares. Out of $40$ events, $33$ were associated with CMEs (eruptive) and $7$ were not associated with CMEs (confined, 2 C-, 2 M- and 3 X-class flares, \citealt{Yashiro2006}). To select our  $40$ events we used  the \verb+RibbonDB+ flare ribbon catalogue that described reconnection flux properties of $3137$ flares  \citep{Kazachenko2017}. 
Our main selection criteria was choosing a sample of events spanning a wide range of AR and reconnection fluxes of flare class M1.0 and above, representative of the medium and strong flares of the \verb+RibbonDB+ catalogue.
 Figure~\ref{db} compares \verb+FlareMagDB+ events with events from the \verb+RibbonDB+: it shows scatter plots of the peak X-ray flux vs. unsigned AR, ribbon and PIL magnetic fluxes for events in both databases. 


For each event in the database, we use 1) a pre-flare HMI/SDO vector magnetogram of $ds_\mathrm{HMI}=0.5''$ spatial resolution, ${\bf B}(x,y)=(B_x,B_y,B_z)(x,y)$, to find magnetic field properties before the flare and 2) a sequence of AIA $1600$\AA{} images, of $dt_\mathrm{1600}=24$s cadence and $ds_\mathrm{1600}=0.''61$ angular resolution, $I_\mathrm{1600}(x,y,t)$, to find areas that have been swept up by flare ribbons during the course of the flare (i.e. cumulative ribbon masks, see \citealt{Kazachenko2017}). For the magnetic field we used a full-disk vector magnetogram data set from the $135$-seconds series (\citealt{Sun2017}, \verb+hmi.B_135S+). To disambiguate the azimuth orientation of the magnetic field we used the radial acute disambiguation method \citep{Hoeksema2014}. We process the UV $1600$\AA{} images in IDL using the \verb+aia_prep.pro+ SolarSoft package and co-align the AIA image sequence in time with the first
frame. We use  \verb+aia_prep.pro+  to align HMI vector magnetic fields maps with the AIA image sequence. In all magnetic field calculations, we set to zero the components of the vector magnetic field ${\bf B}$, where $|B_z|<20G$.  This noise threshold is chosen experimentally to remove areas outside the AR \citep{Hoeksema2014}.
For finding PILs, we first smooth the input magnetic field with a Gaussian function with a window size of $2$ pixels, then isolate areas where vertical magnetic field $|B_z|>200G$ and finally dilate areas of positive and negative flux with a window size of $8$ pixels. This procedure results in PILs that are roughly $8$ AIA pixels or $\approx3500$km wide.
We notice that some of the derived PILs are not directly related to the flaring regions --  there are few false PILs in the isolated-sunspots penumbra and also PILs that lie away from the flares; however the areas and fluxes of these PILs do not contribute significant amounts of flux in our analysis and we thus do not exclude them.
To account for noise in the weak transverse component of {\bf B}, we set $(B_x,B_y)=0$, where $|B_h|<200G$. We define the AR region-of-interest (ROI) as a $800 \times 800$-pixels, or a $390'' \times 390''$, rectangle centered on the AR. We chose the rectangle big enough to include the AR as a whole and small enough to exclude neighboring ARs. We derive the coordinates of the AR center from the Heliophysics Event Catalogue (HEC) maintained by the INAF-Trieste Astronomical Observatory.  

\subsection{Methods: Analyzed physical variables}\label{vars}

We use HMI/SDO photospheric vector magnetic field maps in Cartesian coordinates, ${\bf B}(x,y)=(B_x,B_y,B_z)(x,y)$, to find the following pre-flare magnetic field properties: the potential component of the magnetic field, ${\bf B^\mathrm{p}}=(B_\mathrm{x}^\mathrm{p},B_\mathrm{y}^\mathrm{p},B_\mathrm{z})$, the magnetic shear, $\alpha(x,y)$ \citep{Petrie2019, Wang2017,Gosain2010} and the vertical electric current density, $J_\mathrm{z}(x,y)$ \citep{Liu2017}:
\begin{equation} \label{eqalpha}
    \alpha(x,y) = |B| cos^{-1} \left ( \frac{{\bf B} \cdot {\bf B}^p}{{ B} { B}^p}\right )=|B| \theta, 
\end{equation}
\begin{equation} \label{eqjz}
    J_\mathrm{z}(x,y) = \frac{1}{\mu_0} \nabla \times B_h=\frac{1}{\mu_0} \left( \frac{\partial B_y}{\partial x}-\frac{\partial B_x}{\partial y}\right).
\end{equation}
To calculate the potential field ${\bf B^\mathrm{p}}$, we used the poloidal-toroidal decomposition of the photospheric vector magnetic field (see Appendix A in \citealt{Fisher2010}). 

From the above quantities we then find the following area-integrated physical quantities: the unsigned and the averaged between positive and negative magnetic fluxes ($\Phi_u$ and $\Phi$), the total area ($S$), the mean shear (${\bar \alpha}$), and the unsigned vertical current $( I_\mathrm{z,u})$:
\begin{equation} \label{eqphiu}
    \Phi_u = \int  |B_{n}| \ dS, \;\;\;\;\;\;  \;\;\;\;\;\; S= {\int \limits dS},  
\end{equation}
\begin{equation} \label{eqphi}
     \Phi =\frac{\bigg|\int \limits_\mathrm{B_z>0} B_z dS\bigg| +
    \bigg|\int \limits_\mathrm{B_z<0} B_z dS\bigg|}{2}=
    \frac{|\Phi_{+}|+|\Phi_{-}|}{2},
\end{equation}
\begin{equation} \label{eqalpha1}
    {\bar \alpha} =\frac{\frac{\int \limits_\mathrm{B_z>0} \alpha dS }{S_{+}}+
    \frac{\int \limits_\mathrm{B_z<0} \alpha dS }{S_{-}}}{2}=
    \frac{{\bar \alpha_{+}}+{\bar \alpha_{-}}}{2},
\end{equation}
\begin{equation} \label{eqi}
I_\mathrm{z,u} =\frac{\int \limits_\mathrm{B_z>0} |J_z| dS +
\int \limits_\mathrm{B_z<0} |J_z| dS}{2}=\frac{|I_{z,u+}|+|I_{z,u-}|}{2},
\end{equation}
where $dS$ is the area of integration and ``$+$'' and ``$-$'' subscripts refer to integration over positive and negative polarities, respectively. To describe physical quantities within PIL, ribbon and AR areas we separately integrate over three areas of interest ($dS$): area of the active region ($\verb+AR+$, $|B_z|>20G$), area swept by the flare ribbons  ($\verb+rbn+$, $|B_z|>20G$) and area within the magnetic polarity inversion line ($\verb+PIL+$).
We only consider areas where $|B_h|>200 G$ to calculate $ {\bar \alpha}$ and $I_\mathrm{z,u}$.

In addition to describe the fraction of the AR undergoing magnetic reconnection, we use the percentage of the ribbon-to-AR magnetic fluxes as first described in \citet{Kazachenko2017}:
\begin{equation} \label{eqfrac}
R_\mathrm{\Phi}=\frac{\Phi_\mathrm{ribbon}}{\Phi_\mathrm{AR}}\times100\%.
\end{equation}
 
To estimate the net current we follow the steps below. We first integrate $J_z$ values of different signs in each polarity separately and compute the direct (DC) and return (RC) currents in each polarity \citep{Liu2017,Schmieder2018}. To associate the correct sign of $J_z$ to the DC (and hence to the RC), we find the dominant sign of helicity, $H$, handedness/sign (or $J_z/B_z$) of the AR as a whole.  To define handedness we use orientation of the coronal loops from the coronal images relative to the field orientation of the AR from the magnetogram.
Knowing the correct sign for DC (and RC), we then calculate DC and RC in positive and negative polarities, ($|DC|^+$ and $|RC|^+$) and ($|DC|^-$ and $|RC|^-$), respectively. For right-handed AR with positive helicity ($H>0$):  
\begin{equation}\label{eqdc_rh}
    DC^+=\int \limits_\mathrm{B_z>0,J_z>0} J_{z} \ dS, \;\;\;
    RC^+=\int \limits_\mathrm{B_z>0,J_z<0} J_{z} \ dS,
\end{equation}
\begin{equation}\label{eqrc_rh}
    DC^- =\int \limits_\mathrm{B_z<0,J_z<0} J_{z} \ dS, \;\;\;
    RC^- =\int \limits_\mathrm{B_z<0,J_z>0} J_{z} \ dS,
\end{equation}

For left-handed AR with negative helicity ($H<0$):  
\begin{equation} \label{eqdc_lh}
    DC^+ =\int \limits_\mathrm{B_z>0,J_z<0} J_{z} \ dS, \;\;\;
    RC^+ =\int \limits_\mathrm{B_z>0,J_z>0} J_{z} \ dS,
\end{equation}
\begin{equation} \label{eqrc_lh}
    DC^- =\int \limits_\mathrm{B_z<0,J_z>0} J_{z} \ dS, \;\;\;
    RC^- =\int \limits_\mathrm{B_z<0,J_z<0} J_{z} \ dS.
\end{equation}
From the above direct and return currents in positive and negative polarities, we determine the net current (or an in inverse of neutralization ratio defined in \citealt{Dalmasse2015}):   
\begin{equation} \label{eqcn}
       \bigg|\frac{DC}{RC}\bigg| =\frac{\frac{|DC^+|}{|RC^+|}+ \frac{|DC^-|}{|RC^-|}  }{2}.
\end{equation}
$ \bigg|\frac{DC}{RC}\bigg|$ describes current neutralization within the region of interest, an active region as a whole (AR) or reconnected field lines connecting flare ribbons (ribbon).

To describe CMEs associated with the eruptive events in the \verb+FlareMagDB+, we include CME speeds from \citealt{Yashiro2006} catalog.

\subsection{Methods: Uncertainties}\label{unc}

To estimate the uncertainties in the observed quantities in Eqs.~\ref{eqphi}-\ref{eqcn}, we calculate  their differences within positive (``$+$'') and negative  (``$-$'') magnetic polarities. Since for an isolated active region opposite polarities correspond to the footpoints of a single magnetic flux system, estimates within opposite polarities should be equal within the observational errors.\footnote{We note that our shear parameter does not obey any conservation principle. Also, our PIL algorithm does not enforce balance in either flux or current. Nonetheless, we do not expect substantial imbalances in these quantities, so it is plausible to use excursions from zero in these quantities to indicate large uncertainties, perhaps arising from systematic effects.} Adopting this hypothesis, we define the error proxies for signed magnetic fluxes, mean shears, total vertical currents and the net currents in the following way:
 \begin{equation} \label{eqdphi}
    \Delta \Phi = \frac{|\Phi_+| - |\Phi_-|}{2}, \;\;\;\;\;\;\;\;\;\;  \Delta \bar \alpha=  \frac{|\bar\alpha_{+}|- |\bar\alpha_{-}|}{2},\;\;\;\;\;\;\;\;\;\;     
\end{equation}
\begin{equation} \label{eqdlz}
    \Delta I_z = \frac{|I_{z,u+}| - |I_{z,u-}|}{2}, \;\;\;\;\;\;\;\;\;\;  \Delta \bigg|\frac{DC}{RC}\bigg|= \frac{\frac{|DC^+|}{|RC^+|}- \frac{|DC^-|}{|RC^-|}  }{2}. \;\;\;\;\;\;\;\;\;\;     
\end{equation}
From the way we construct these metrics we expect that for ideal measurements these error proxies should be zero.

\subsection{Statistical Analysis}\label{stat}

To quantitatively describe the relationship between different properties of flares and ARs,
e.g. $\mathbb{X}$ and $\mathbb{Y}$, we use Spearman ranking correlation
coefficient, $r_s(\mathbb{X},\mathbb{Y})$ (see Figure~\ref{heatmap}). Unlike the Pearson correlation coefficient that is used to measure {\it linear} relationship between variables---and therefore is not optimal for
non-linearly related variables---the Spearman rank correlation provides a measure of a {\it monotonic}
relationship between variables. 
We describe the qualitative strength of the correlation using
the following guide for the absolute value of $r_s$ \citep{Kazachenko2017}:
$r_s \in [0.2,0.39]$ -- weak,
$r_s \in [0.4,0.59]$ -- moderate,
$r_s \in [0.6,0.79]$ -- strong, and
$r_s \in [0.8,1.0]$ -- very strong.
When the correlation coefficient is moderate or greater ($r_s(\mathbb{X},\mathbb{Y})>0.4$), we
fit the relationship between $\mathbb{X}$ and $\mathbb{Y}$ with a power-law function
\begin{equation}
\mathbb{Y} = a \mathbb{X}^b \ .
\end{equation}
We use Levenberg-Marquardt non-linear least-squares minimization method to find the scaling factor $a$ and exponent $b$.
\section{Results}\label{results}    

\section{Results: Observational analysis of the dataset}\label{res01}
In this section we first show results of the magnetic field analysis within PIL, ribbon and AR areas for an example AR 12158 (\S\ref{example}, Figure~\ref{db_example}).  We then describe results of the statistical analysis for $40$ events of the \verb+FlareMagDB+: magnetic flux (\S\ref{mag}),  flux ratio (\S\ref{frac}), magnetic shear (\S\ref{shear}) and total and net currents (\S\ref{dcrc}). Table~\ref{events} contains the list of all the variables for $40$ events. Table~\ref{summary_table} and Figure~\ref{heatmap}  show variables' typical range and correlation matrix between $18$ \verb+FlareMagDB+ variables.

\subsection{FlareMagDB example event: X1.6 flare in AR 12158}\label{example}
\begin{figure*}[htb!] \centering
\resizebox{1.0\hsize}{!}{\includegraphics[angle=0]{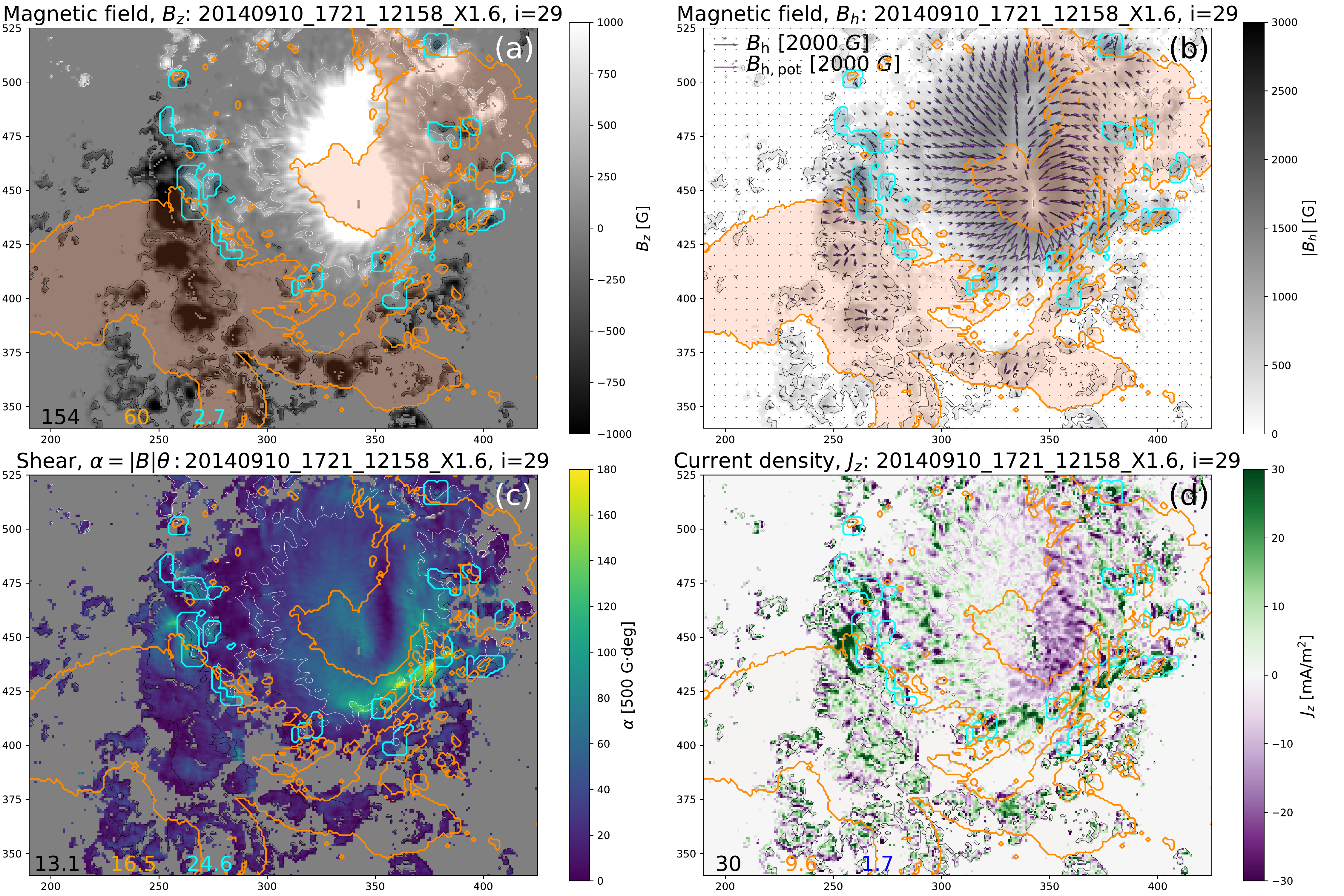}} \caption{Example of an X1.6 flare on September 10 2014 in AR 12158 from the FlareMagDB database: clockwise from upper left, vertical and horizontal components of the magnetic field, magnetic shear and the vertical current density before the flare. Orange and blue contours outline ribbon and PIL areas, respectively. Black and blue arrows show observed and potential components of the horizontal magnetic field. See \S\ref{example}.}  
\label{db_example}
\end{figure*}

Figure~\ref{db_example}  shows an example of our analysis
for one event in the \verb+FlareMagDB+ database, an X1.6 flare on September 10 2014 in AR 12158 (event number $i=29$ in Table~\ref{events}). Four panels show vertical and horizontal magnetic field components, shear and current density maps right before the flare. Orange contours indicate areas swept by the flare ribbon. Cyan contours show locations of the PILs.  In Table~\ref{events} we show the results of the quantitative analysis for this event (marked with ``+'') and all the other  \verb+FlareMagDB+ events.
From the magnetic flux analysis we find that
 the AR magnetic flux is $\Phi_\mathrm{AR}=154\times10^{20}Mx$, the magnetic flux swept by flare ribbons is $\Phi_\mathrm{rbn}=60\times10^{20}Mx$, and 
 the PIL magnetic flux is $\Phi_\mathrm{PIL}=2.7\times10^{20}Mx$ (Figure~\ref{db_example}a). 
We find that the mean shear is largest within the PIL area gradually decreasing within the ribbon and AR areas ($\bar \alpha_\mathrm{PIL}=24.6\times10^3 G\cdot deg$ vs. $\bar \alpha_\mathrm{ribbon}=16.5\times10^3 G\cdot deg$  and $\bar \alpha_\mathrm{AR}=13.1\times10^3 G\cdot deg$), which corresponds to field being increasingly sheared as we go from AR to ribbon and PIL areas. In \ref{db_example}(c) we can see this transition from the most yellow, or largest-shear areas, near the cyan PILs to bluer, i.e smaller-shear areas, outside the ribbon and PIL contours in the AR as a whole  (Figure~\ref{db_example}b,c).
As for the net vertical current, that describes current neutralization, we find that the current is non-neutralized within separate polarities, increasing from AR to PIL areas: $|DC/RC|[\mathrm{AR},\mathrm{rbn},\mathrm{PIL}]=[1.2,1.6,1.6]$.
In \ref{db_example}d we see this current imbalance as preferred current colors in opposite magnetic polarities -- notice how current is more violet within positive ribbon (orange contour in the top right) and more green within negative ribbon (bottom left).

\begin{table*} \caption{FlareMagDB database: magnetic field properties of $40$ flares within AR, ribbon and PIL areas.  $H$, $\Phi$, $\bar \alpha$, $I_\mathrm{z,u}$, $|DC/RC|$ and $V_\mathrm{CME}$ refer to AR handedness, magnetic flux, mean magnetic shear, total, net currents and CME speed, respectively. $\bigtriangleup$s mark confined events. ``+' marks an example event (\S\ref{example}). See \S\ref{res01} for details. }\small \begin{center}\begin{tabular}{lcccr|ccc|ccc|ccc|ccc|l}
 \toprule $i$ & $T_\mathrm{start}$ & Flare & AR & H & \multicolumn{3}{c|}{$\Phi, 10^{20} Mx$} &   \multicolumn{3}{c|}{$\bar \alpha , 10^{3} G\cdot deg$} &\multicolumn{3}{c|}{$I_\mathrm{z,u}, 10^{12} A$} &\multicolumn{3}{c|}{$|DC/RC|$} & $V_\mathrm{CME}$ \\  & & class & number & +/- & AR & rbn & PIL &  AR & rbn & PIL & AR & rbn & PIL & AR & rbn & PIL & $km/s$ \\\hline \hline00 & 2010-08-07 17:55 & M1.0 & 11093 & -1&64 & 12 & 0.1 & 8.4 & 12.0 & 11.0 & 45 & 4.8 & 0.0 & 1.0 & 1.3 & 1.9 & 871 \\01 & 2011-02-15 01:44 & X2.2 & 11158 & 1&139 & 53 & 5.5 & 12.8 & 25.1 & 51.5 & 49 & 13.8 & 3.6 & 1.1 & 1.5 & 1.7 & 669 \\02 & 2011-08-03 13:17 & M6.0 & 11261 & 1&127 & 23 & 7.2 & 11.9 & 18.5 & 34.5 & 61 & 9.5 & 5.7 & 1.2 & 1.6 & 4.2 & 610 \\03 & 2011-09-06 22:12 & X2.1 & 11283 & 1&122 & 23 & 1.3 & 7.8 & 20.9 & 41.7 & 48 & 7.7 & 0.8 & 1.1 & 2.0 & 3.2 & 575 \\04 & 2011-10-02 00:37 & M3.9 & 11305 & -1&87 & 10 & 2.3 & 7.7 & 16.3 & 34.9 & 37 & 3.6 & 1.5 & 1.2 & 2.0 & 2.8 & 259 \\\hline05 & 2011-11-15 12:30 & M1.9 & 11346 & -1&67 & 06 & 1.1 & 7.4 & 15.1 & 19.2 & 123 & 5.5 & 1.1 & 1 & 1.8 & 1.9 & 300 \\06$^\bigtriangleup$ & 2011-12-27 04:11 & C8.9 & 11386 & 1&74 & 04 & 0.7 & 7.7 & 10.5 & 13.1 & 170 & 3 & 0.8 & 1.1 & 1.0 & 1.6 & - \\07 & 2012-01-19 13:44 & M3.2 & 11402 & -1&215 & 10 & 2.9 & 10.7 & 17.9 & 21.8 & 230 & 5.0 & 2.6 & 1.0 & 1 & 1.2 & 1120 \\08 & 2012-01-23 03:38 & M8.7 & 11402 & -1&222 & 35 & 3.5 & 10.7 & 16.9 & 21.2 & 160 & 18.2 & 3.3 & 1.0 & 1.1 & 2.1 & 2175 \\09 & 2012-03-07 00:02 & X5.4 & 11429 & -1&188 & 76 & 12.4 & 14.3 & 29.3 & 41.3 & 124 & 27.2 & 7.7 & 1.1 & 1.6 & 2.0 & 2684 \\\hline10 & 2012-03-09 03:22 & M6.3 & 11429 & -1&212 & 59 & 5.6 & 12.3 & 19.1 & 43.1 & 99 & 14.7 & 3.4 & 1.2 & 1.8 & 4.4 & 950 \\11 & 2012-03-10 17:15 & M8.4 & 11429 & -1&184 & 58 & 3.0 & 12.9 & 18.8 & 32.5 & 87 & 19.0 & 2.4 & 1.2 & 1.4 & 2.3 & 1379 \\12 & 2012-03-14 15:08 & M2.8 & 11432 & 1&74 & 12 & 0.3 & 7.9 & 10.8 & 9.1 & 30 & 3.5 & 0.2 & 1.0 & 1.1 & 0.7 & 411 \\13 & 2012-07-12 15:37 & X1.4 & 11520 & 1&356 & 40 & 7.6 & 15.5 & 22.1 & 29.4 & 94 & 6.6 & 3.9 & 1.2 & 1.7 & 1.9 & 885 \\14 & 2012-11-21 06:45 & M1.4 & 11618 & -1&132 & 15 & 4.9 & 11.4 & 23.5 & 30.0 & 37 & 4.3 & 3 & 1.2 & 2.3 & 2.4 & 410 \\\hline15 & 2013-04-11 06:55 & M6.5 & 11719 & -1&123 & 19 & 0.2 & 8.3 & 12.6 & 11.7 & 52 & 4.9 & 0.2 & 1.0 & 1.1 & 1.8 & 861 \\16$^\bigtriangleup$ & 2013-05-16 21:36 & M1.3 & 11748 & -1&69 & 07 & 1.8 & 8.9 & 16.9 & 25.0 & 85 & 5.0 & 1.5 & 1.0 & 1.8 & 2.5 & - \\17 & 2013-05-31 19:52 & M1.0 & 11760 & -1&37 & 03 & 0.3 & 7.0 & 16.4 & 37.6 & 95 & 2.8 & 0.3 & 1.0 & 1.3 & 2.7 & 388 \\18 & 2013-08-17 18:49 & M1.4 & 11818 & 1&108 & 20 & 2.2 & 10.1 & 24.1 & 61.9 & 89 & 6.3 & 1.2 & 1.1 & 1.9 & 4.9 & 1202 \\19$^\bigtriangleup$ & 2013-12-28 17:53 & C9.3 & 11936 & -1&112 & 06 & 1.2 & 8.2 & 17.9 & 25.5 & 37 & 2.1 & 0.7 & 1.0 & 2.8 & 2.8 & - \\\hline20 & 2014-01-07 18:04 & X1.2 & 11944 & 1&355 & 50 & 3.7 & 9.6 & 10.1 & 21.8 & 77 & 9.4 & 2.2 & 1.1 & 1.1 & 1.7 & 1830 \\21 & 2014-01-31 15:32 & M1.1 & 11968 & -1&105 & 07 & 0.3 & 8.3 & 11.3 & 8.8 & 123 & 5.5 & 0.3 & 1.0 & 0.9 & 1.7 & 462 \\22 & 2014-02-01 07:14 & M3.0 & 11967 & -1&336 & 23 & 9.6 & 14.4 & 23.6 & 34.4 & 133 & 7.2 & 7.1 & 1.1 & 1.7 & 1.0 & - \\23 & 2014-02-12 03:52 & M3.7 & 11974 & -1&196 & 32 & 6.8 & 8.6 & 17.0 & 24.9 & 69 & 11.8 & 4.5 & 1.0 & 1.3 & 1.2 & 373 \\24 & 2014-03-20 03:42 & M1.7 & 12010 & 1&103 & 07 & 1.6 & 8.1 & 14.5 & 15.0 & 83 & 3.5 & 1.5 & 1.0 & 0.8 & 1.9 & 740 \\\hline25 & 2014-08-01 17:55 & M1.5 & 12127 & 1&162 & 21 & 0.6 & 9.3 & 12.5 & 12.0 & 69 & 6.0 & 0.5 & 1.0 & 1 & 1.1 & 789 \\26 & 2014-08-25 14:46 & M2.0 & 12146 & -1&114 & 13 & 2.4 & 11.1 & 25.9 & 38.2 & 59 & 5.9 & 1.7 & 1.1 & 2.4 & 1.5 & 555 \\27 & 2014-08-25 20:06 & M3.9 & 12146 & -1&106 & 13 & 2.2 & 9.8 & 28.0 & 35.0 & 74 & 6.1 & 1.9 & 1.1 & 2.0 & 2.0 & 711 \\28 & 2014-09-08 23:12 & M4.5 & 12158 & -1&140 & 28 & 3.7 & 13.1 & 21.5 & 25.6 & 65 & 6.3 & 2.8 & 1.2 & 1.4 & 1.3 & 920 \\29$^+$ & 2014-09-10 17:21 & X1.6 & 12158 & -1&154 & 60 & 2.7 & 13.1 & 16.5 & 24.6 & 30 & 9.6 & 1.7 & 1.2 & 1.6 & 1.6 & 1267 \\\hline30 & 2014-09-28 02:39 & M5.1 & 12173 & -1&262 & 21 & 2.9 & 10.4 & 17.5 & 21.2 & 137 & 8.0 & 2.8 & 1.0 & 1.8 & 1.0 & 215 \\31$^\bigtriangleup$ & 2014-10-22 14:02 & X1.6 & 12192 & -1&614 & 70 & 15.9 & 13.2 & 21.3 & 20.2 & 192 & 17.4 & 10.8 & 1.0 & 1.2 & 0.9 & - \\32$^\bigtriangleup$ & 2014-10-24 21:07 & X3.1 & 12192 & -1&661 & 102 & 12.5 & 13.2 & 18.2 & 19.5 & 209 & 29.3 & 9.2 & 1.0 & 1.1 & 0.8 & - \\33$^\bigtriangleup$ & 2014-10-25 16:55 & X1.0 & 12192 & -1&632 & 77 & 14.8 & 13.0 & 21.5 & 20.1 & 350 & 24.0 & 11.0 & 1.0 & 1.2 & 0.8 & - \\34$^\bigtriangleup$ & 2014-12-17 00:57 & M1.5 & 12242 & 1&273 & 10 & 2.4 & 11.1 & 20.5 & 24.4 & 147 & 3.4 & 2.0 & 1.0 & 1.8 & 1.0 & - \\\hline35 & 2014-12-17 04:25 & M8.7 & 12242 & 1&290 & 34 & 2.5 & 12.5 & 18.1 & 27.4 & 89 & 6.1 & 2.0 & 1.1 & 1.9 & 0.9 & 587 \\36 & 2014-12-18 21:41 & M6.9 & 12241 & 1&227 & 37 & 4.4 & 9.9 & 18.2 & 33.4 & 62 & 9.0 & 2.8 & 1.2 & 2.6 & 1.3 & 1195 \\37 & 2014-12-20 00:11 & X1.8 & 12242 & 1&354 & 80 & 7.1 & 11.5 & 19.9 & 25.5 & 161 & 30.5 & 5.1 & 1.1 & 1.2 & 1.7 & 841 \\38 & 2015-11-04 13:31 & M3.7 & 12443 & 1&206 & 33 & 0.9 & 7.0 & 7.8 & 20.4 & 40 & 6.1 & 0.7 & 1.0 & 1.2 & 0.7 & 578 \\39 & 2015-11-09 12:49 & M3.9 & 12449 & 1&73 & 17 & 0.7 & 8.4 & 15.6 & 11.3 & 125 & 9.5 & 0.5 & 1.0 & 0.9 & 0.9 & 1041 \\\hline\hline $i$ & $T_\mathrm{start}$ & Flare & AR & H & \multicolumn{3}{c|}{$\Phi, 10^{20} Mx$} &   \multicolumn{3}{c|}{$\bar \alpha , 10^{3} G\cdot deg$} &\multicolumn{3}{c|}{$I_\mathrm{z,u}, 10^{12} A$} &\multicolumn{3}{c}{$|DC/RC|$}  & $V_\mathrm{CME}$ \\ \bottomrule \end{tabular}\normalsize \label{events} \end{center} \end{table*}

\begin{table*} \caption{Typical range of active-region, flare-ribbon and PIL properties, $\mathbb{X}$, for $40$ events of FlareMagDB database and ARMS simulation. $\mathbb{X}$ is either magnetic flux $\Phi$, ribbon-to-AR fractions of magnetic flux $R_\mathrm{\Phi}$, mean magnetic shear $\bar\alpha$, total unsigned current $I_\mathrm{z,u}$, or net current $|DC/RC|$. We describe the typical range of each quantity as the $20^{th}$ to $80^{th}$ percentile, $\mathbb{X}[P_{20},P_{80}]$.  For more details see Figures~\ref{4by4}-\ref{2by2}.}\small \begin{center}\begin{tabular}{lccc ccc}  
\toprule
  & \multicolumn{3}{c}{\textbf{OBSERVATIONS (FlareMagDB)}}   & \multicolumn{3}{c}{\textbf{SIMULATIONS (ARMS)}} \\
   \cmidrule(r){2-4}  \cmidrule(r){5-7}
Quantity &  \multicolumn{3}{c}{Typical range,  $\mathbb{X}[P_{20},P_{80}]$}  & \multicolumn{3}{c}{$\mathbb{X}_\mathrm{ARMS}$} \\
 $\mathbb{X}$ & AR & Ribbon & PIL &  AR & Ribbon & PIL \\
\midrule
Magnetic Flux, $\Phi~[10^{20} Mx] $ & $[100,276]$ & [10,54] & [0.9,6.9] & 41& 18 & 0.3 \\
Reconnection flux fraction,  $R_\Phi~[\%]$ &-&  [8.3,20.1] & -&  -& 43 &- \\
Mean Magnetic Shear,  $\bar \alpha~[10^3 G\cdot deg]$& [8.2,12.9]& [14.2,21.7] & [18.3,34.9]& 0.06& 0.13 & 0.17\\
Total Current, $I_\mathrm{z,u}~[10^{12} A]$ &  [50,139]& [4.7,14]& [0.7,4.0] & 0.3& 0.2 & 0.003\\
Net Current, $|DC/RC|$~[-] & [1.0,1.16] & [1.1,1.9]& [1.0,2.4] & 4.9& 10.5 & 3.1\\
\bottomrule
\end{tabular}
\normalsize
\label{summary_table}
\end{center}
\end{table*}

\begin{figure*}
    \includegraphics[angle=0,trim=3.6cm 6cm 6.5cm 0cm,clip,width=0.5\linewidth]{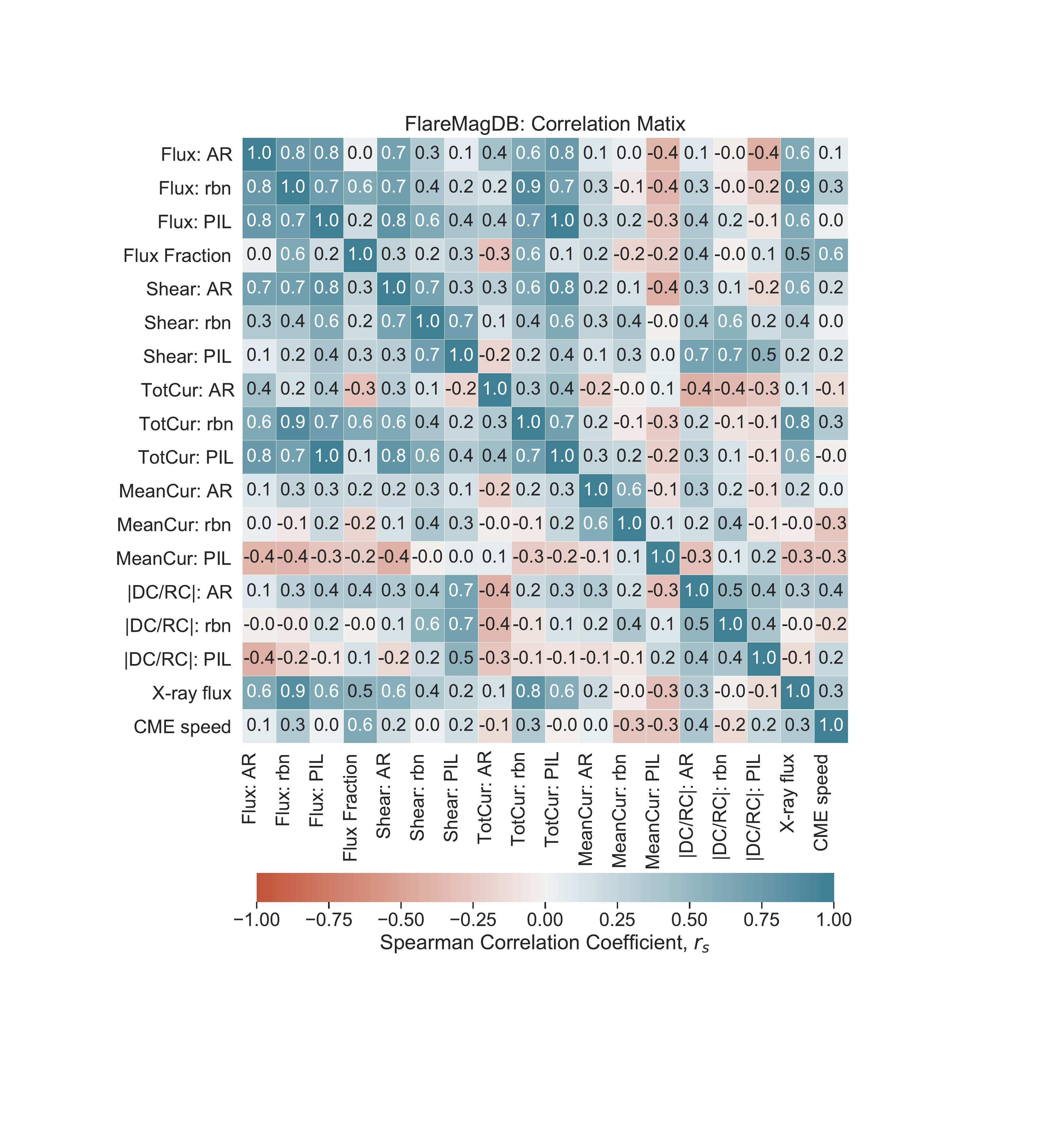}
    \caption{{\it FlareMagDB Correlation matrix:} Spearman correlation coefficients, $r_s$, between variables of FlareMagDB,  $\mathbb{X}$. Here $\mathbb{X}$ is either magnetic flux, ribbon-to-AR fraction of magnetic flux, mean magnetic shear, total current, mean current, net current within AR, ribbon and PIL areas, GOES peak X-ray flux or CME speed. Colors correspond to the strength of the correlation coefficient, $r_s$, between each variable pair. We verbally describe the strength of the correlation using the following guide for the absolute value of $r_s$: $r_s=[0.2,0.39]$ -- weak,  $r_s=[0.4,0.59]$ -- moderate, $r_s=[0.6,0.79]$ -- strong and $r_s=[0.8,1.0]$ -- very strong correlation.}\label{heatmap}
\end{figure*}


\subsection{FlareMagDB Statistics: Magnetic flux}\label{mag}
Figure~\ref{bz_examples} shows vertical magnetic field maps, $B_z(x,y)$, of $40$ events in the \verb+FlareMagDB+ database. Does the geometry of PIL affect where flare ribbons of the next flare would be? Does a stronger PIL necessarily mean that the flare ribbons of the next flare would lie close to PIL? According to the standard flare model, ribbons should never cross PILs and should lie on both sides of the PIL at a distance defined by the reconnection process and the structure of the pre-flare coronal magnetic field. We examine spatial distribution of cyan and orange contours showing PIL and ribbon areas (see Figure~\ref{bz_examples}). We find that ribbons and PIL-masks generally sample different parts of the ARs. For example, for the $10$ strongest flares, X-class and above, only $3$ X-class flares have ribbons lying close to strong PIL areas ($i=1, 3,  9$). In the other $7$ X-class flares, ribbons lie away from strong PIL areas  ($i=13, 20, 29, 31, 32, 33, 37$).   From this comparison we conclude that while large amounts of near-PIL flux indicate a large flare is more likely, the morphology of PILs has almost {\it no bearing} on the morphology of flare ribbons.
We then use these $B_z(x,y)$ maps to find unsigned and averaged between opposite polarities vertical magnetic fluxes, $\Phi_u$ and $\Phi$ (Eq.~\ref{eqphiu}-\ref{eqphi}), for each event within the AR, ribbon and PIL areas (see contours).

Figure~\ref{db} shows scatter plots of pre-flare {\it unsigned} magnetic fluxes, $\Phi_u$ (see Eq.~\ref{eqphiu}), within AR, reconnection and PIL area vs. peak X-ray flux for each event in the \verb+FlareMagDB+ database. Rainbow color corresponds to events from the $3137$-flares \verb+RibbonDB+ catalogue ($n=3137$, \citealt{Kazachenko2017}). Red color shows a subset of $40$ \verb+FlareMagDB+ events selected in this paper. For events from the \verb+FlareMagDB+  we find a moderate correlation between the unsigned AR flux and the flare peak X-ray flux. This correlation is stronger than the correlation found for \verb+RibbonDB+: Spearman correlation coefficient $r_s=0.5$ for $n=40$ vs. $r_s=0.2$ for $n=3137$ (left plot). We also find a strong correlation between the unsigned reconnection flux and the flare peak X-ray flux: $r_s=0.8$ for $n=40$ vs. $r_s=0.7$ for $n=3137$ (middle plot).  We explain the difference in correlation coefficients between a subsample of $n=40$ events and the full sample of $n=3137$ by the high heterogeneity of the full \verb+RibbonDB+ dataset. We also note that  using $H_\alpha$ dataset from the Kanzelhohe Observatory instead of the AIA/SDO observations, \citealt{Tschernitz2018} found stronger correlations ($r_s=0.9$) in agreement with our findings here. Finally, we find a strong correlation between the PIL vertical magnetic flux and the flare peak X-ray flux: $r_s=0.6$ (right plot).

\begin{figure*}[htb!]
  \centering
 \resizebox{1.0\hsize}{!}{\includegraphics[angle=0,trim=3.5cm 1cm 4cm 3cm,clip,width=\textwidth]{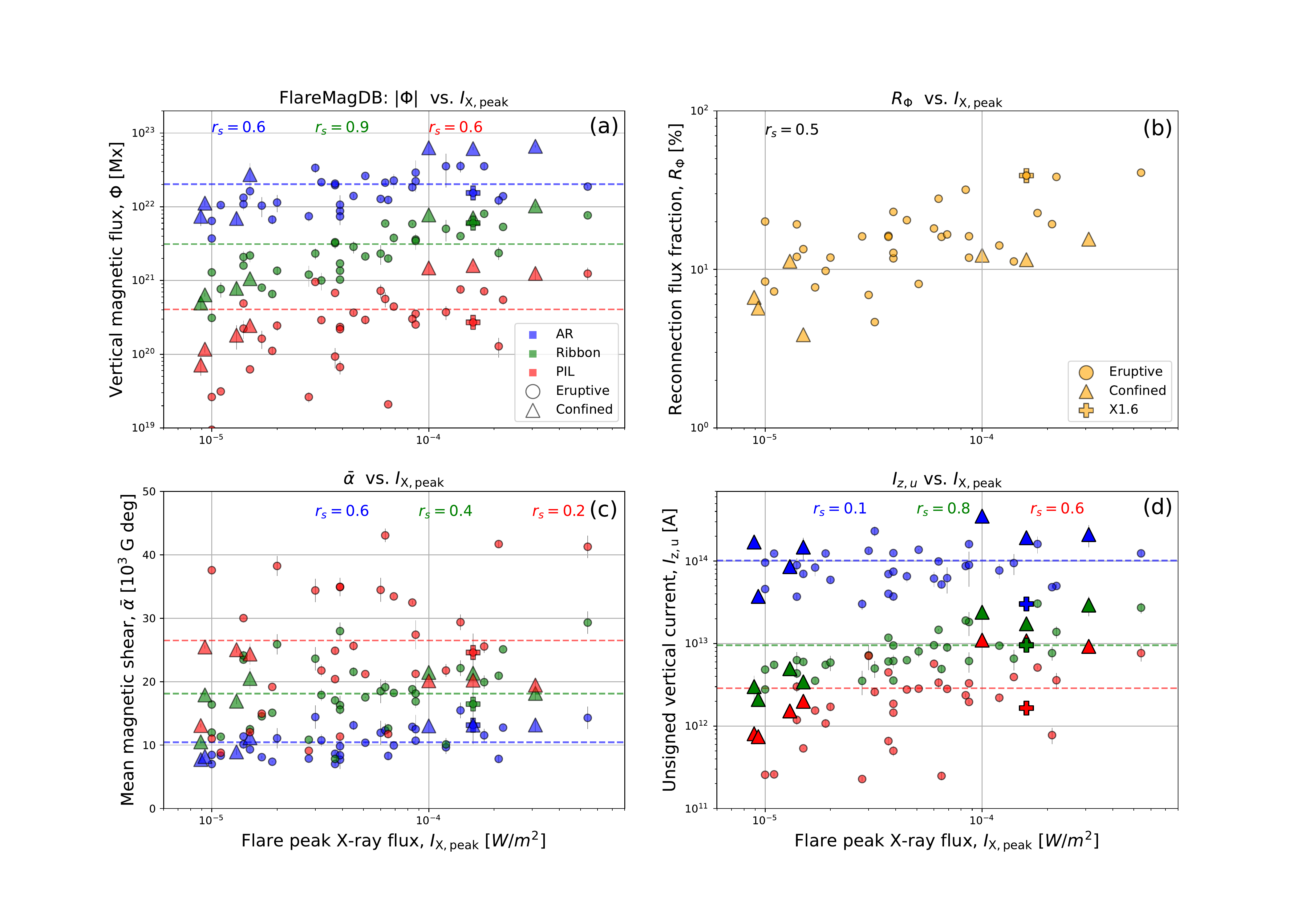}}
  \caption{{\it FlareMagDB results}:  magnetic field properties for $40$ events of FlareMagDB database. Vertical magnetic flux ({\it top left}, \S\ref{mag}), reconnection flux fraction ({\it top right}, \S\ref{frac}), mean magnetic shear ({\it bottom left}, \S\ref{shear}) and the total unsigned vertical current ({\it bottom right}, \S\ref{dcrc}) vs. the peak X-ray flux within AR, ribbon and PIL areas. The error bars correspond to error proxies as defined in \S\ref{unc}.}
  \label{4by4}
\end{figure*}

In Figure~\ref{4by4}a we show AR, ribbon and PIL magnetic fluxes averaged between opposite polarities, $\Phi$  vs. the peak X-ray flux for the \verb+FlareMagDB+ alone (see Eq.~\ref{eqphi}). 
 By definition, these averaged fluxes are around $2$-times smaller than unsigned fluxes from Figure~\ref{db}.
We find that the flare size is moderately correlated with the AR magnetic flux and the strength of the PIL and very strongly correlated with the amount of reconnecting magnetic flux. In Table~\ref{summary_table} we describe the typical range of AR, ribbon and PIL fluxes consistent with earlier results \citep{Kazachenko2017}.


In Figure ~\ref{heatmap} we show the Spearman correlation coefficients between AR, ribbon and PIL fluxes and other \verb+FlareMagDB+ variables, the correlation matrix. We find very strong correlations ($r_s>0.8$) between magnetic flux and total current within PIL.
Note that PIL and AR fluxes have weaker correlation with the Peak X-ray flux than the ribbon reconnection flux. We discuss these relationships further in \S\ref{disc}.
\subsection{FlareMagDB Statistics: Reconnection Flux Fraction}\label{frac}

In Figure~\ref{4by4}b we show the fraction of the magnetic flux participating in the flare, $R_\mathrm{\Phi}$ (see Eq~\ref{eqfrac}) vs. the peak X-ray flux. We find that $R_\mathrm{\Phi}$ has a moderate correlation with the peak X-ray flux ($r_s=0.5$).  The reconnection flux fraction has the typical range of $8.3\%$ to $20.1\%$, consistent with previous results \citep{Kazachenko2017}.


\subsection{FlareMagDB Statistics: Magnetic Shear}\label{shear}
Figure~\ref{shear_examples} in Appendix~\ref{app01} shows magnetic shear maps, $\alpha(x,y)$, for all events of the database. From these maps we find that the spatial structure of the magnetic shear highly varies for different events.  To understand these variations in a quantitative way, we use individual shear maps to find the mean magnetic shear, $\bar \alpha$, within the  AR as a whole, ribbon and PIL areas (see Eq.~\ref{eqalpha}).

Figure~\ref{4by4}c shows the mean magnetic shear vs. the peak X-ray flux.  We find that the mean magnetic shear in AR, ribbon and PIL areas is $[10,18,26]\times10^3 G\cdot deg$ with the  $20^{th}$ to $80^{th}$ percentile of $[8.2,12.9]\times10^3 G\cdot deg$, $[14.2,21.7]\times10^3 G\cdot deg$ and $[18.3,34.9]\times10^3 G\cdot deg$, respectively (see Table~\ref{summary_table}). In other words, as we go from the AR as a whole to ribbon and PIL areas, the magnetic field becomes stronger and more sheared. In addition, we find that the peak X-ray flux has a strong correlation with the mean magnetic shear within the AR and weak to moderate correlations with the mean magnetic shear within PIL and ribbon areas.  

 
%
\subsection{FlareMagDB Statistics: Vertical Current and Net Current}\label{dcrc}

Figure~\ref{jz_examples} in Appendix~\ref{app01} shows vertical current density maps, $J_z(x,y)$, for all events of the database. Zooming in into individual $J_z$ images, we see that the vertical current density consists of long and short structures, or ``threads'' and  ``patches'', of violet and green colors that correspond to positive and negative vertical currents. While we find elongated thread structures in some events, e.g. events $i=1$ (\verb+01_20110215_0144_11158_X2.2+), $i=22$ (\verb+22_20140201_0714_11967_M3.0+) or $i=36$ (\verb+36_20141218_2141_12241_M6.9+),  from our limited sample we do not find any relationship between the presence of these structures and occurrence of large flares or CMEs (see e.g. a non-thread-like event \verb+29_2014_0910_1721_12158_X1.6+, where an X1.6 occurred with a CME).
 Instead we find that the current density maps have all kinds of shapes of varying size that are not related to flare size or flare/CME productivity.

To quantify the global properties of $J_z$, we use $J_z$ maps to compute the total unsigned vertical current, $I_\mathrm{z,u}$, direct current (DC), return current (RC) and their ratio, or net current, $|DC/RC|$  as described in Eqns.~\ref{eqi}-\ref{eqcn}.

Figure~\ref{4by4}d shows the scatter plot between the total unsigned vertical current and the peak X-ray flux before the flare within PIL, ribbon and AR areas. We find that the total unsigned vertical current is largest within AR areas decreasing gradually within ribbon and PIL areas. This relationship is not surprising and reflects the decreasing area of integration as we go from AR to ribbon and PIL areas. Comparing Figure~\ref{4by4}a and Figure~\ref{4by4}d, i.e. the vertical-magnetic-flux and the total-vertical-current scatter plots, we see similar scattering with the peak X-ray flux. The Spearman correlation coefficient between the peak X-ray flux and the total vertical current ranges from weak for ARs to strong for ribbon and PIL areas.  

\begin{figure*}[tb!]
  \centering
 \resizebox{0.63\hsize}{!}{\includegraphics[angle=0,trim=2cm 0cm 3cm 0cm,clip,width=\textwidth]{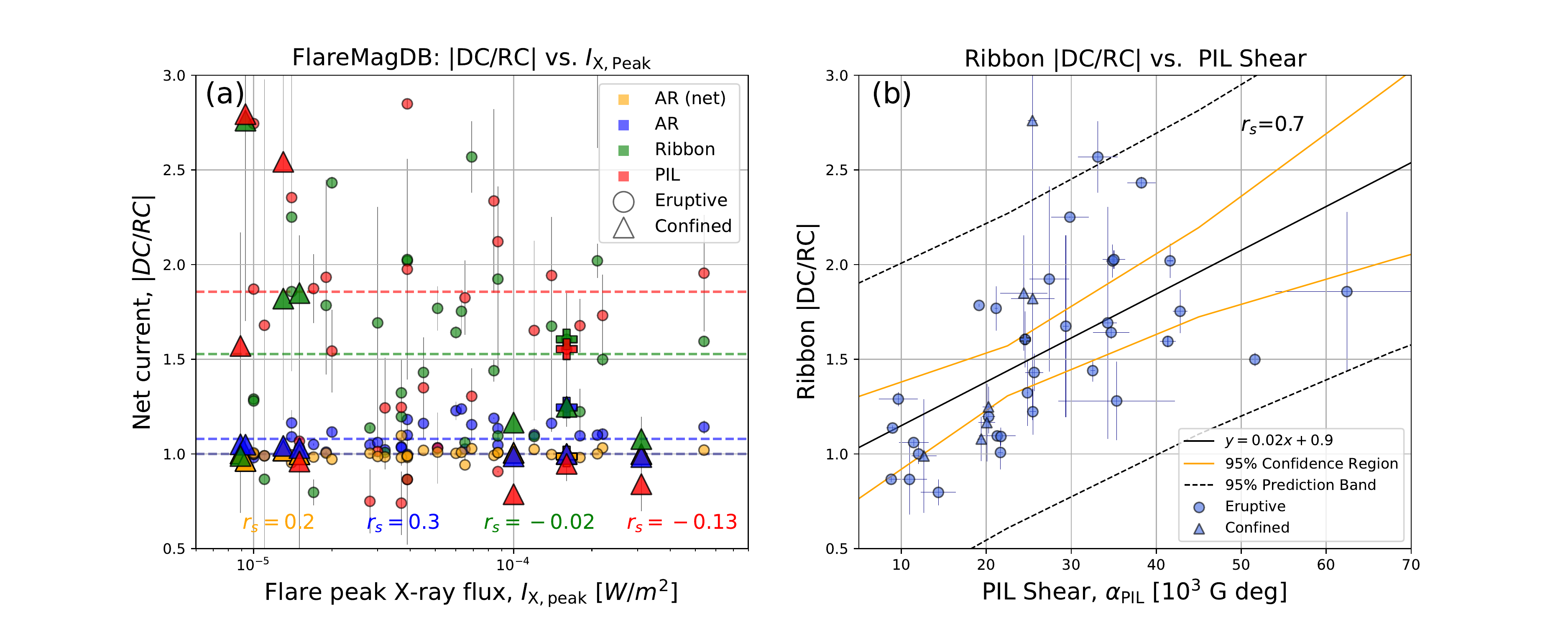}}
  \resizebox{0.36\hsize}{!}{\includegraphics[angle=0,trim=3cm 0cm 6cm 0cm,clip,width=\textwidth]{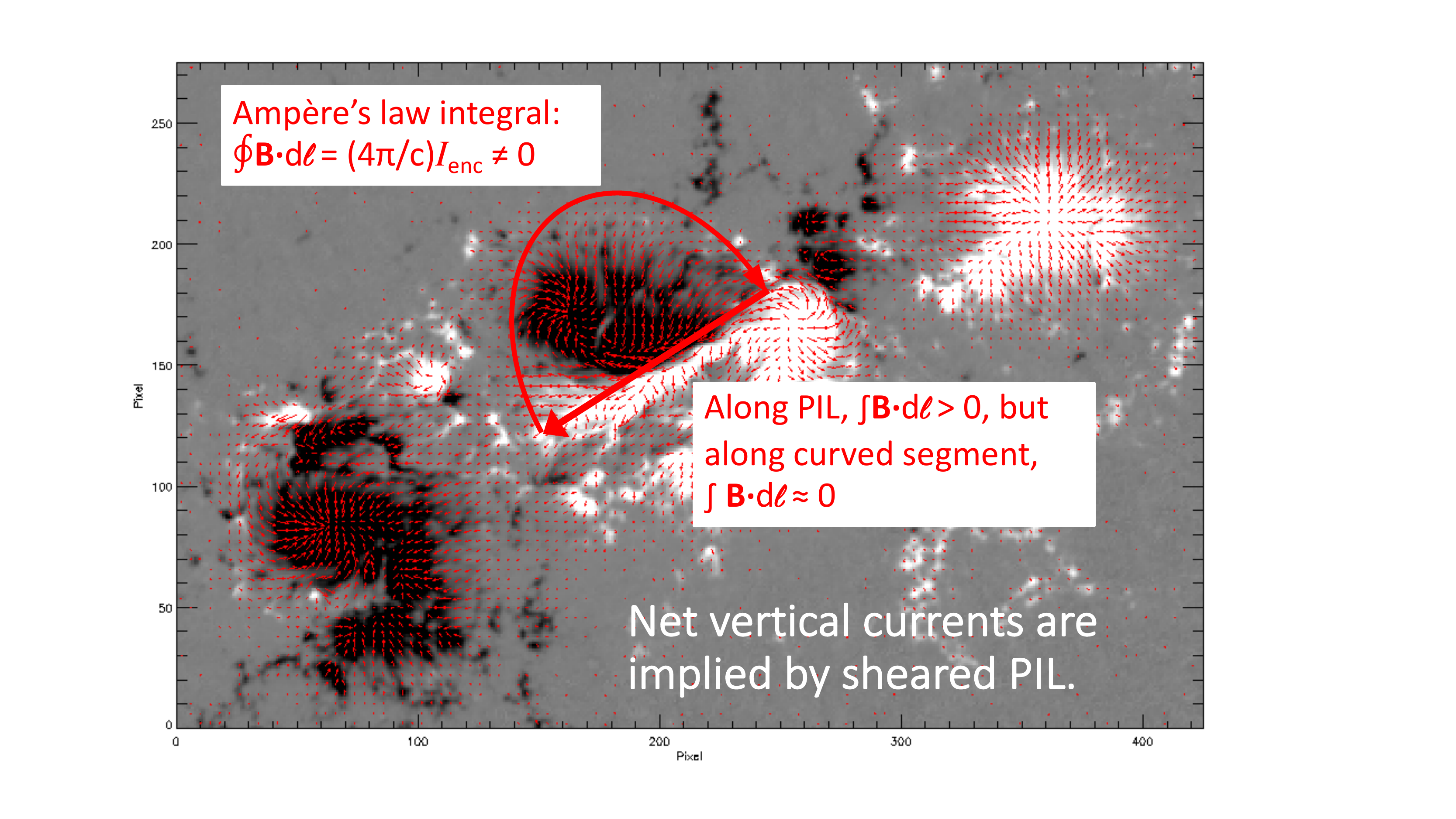}}
  \caption{{\it FlareMagDB results}: net current properties for $40$ events of FlareMagDB database. {\it Left panel, (a):} DC/RC vs. the peak X-ray flux; {\it Middle panel, (b):} DC/RC within ribbons vs. mean PIL shear angle. The error bars correspond to error proxies as defined in \S\ref{unc}.
 {\it Right panel}: explanation of the scaling law in Panel (b), between the net current and shear at PIL, as a consequence of the Amp\`ere's Law. See \S\ref{dcrc}.}
  \label{2by2}
\end{figure*}

In Figure~\ref{2by2} we describe the imbalance of the vertical current within each magnetic polarity,  {\it the net current}, in AR, ribbon and PIL areas and compare these to various magnetic field properties. Our objective here is to understand the cause of the net current. In Figure~\ref{2by2}a we compare the  $|DC/RC|$ with the peak X-ray flux. We find that $|DC/RC|$ and the flare peak X-ray flux have weak correlation, implying that the net current is not related to how large the next flare might be. We find that the net current is larger within PILs, than within ribbons and separate AR polarities ( $\overline{|DC/RC|}=[1.1,1.5,1.7]$ within ARs, ribbons and PILs, respectively), i.e. the current is more non-neutralized within ribbons' than within AR's polarities. If we look not at the separate polarities but at the AR as a whole, we find that $|DC/RC|\approx1$, i.e the current is neutralized.

To understand the cause of the net current, in Figure~\ref{2by2}b we compare the pre-flare mean shear with the net current within flare ribbons. We find a strong correlation between the net current within ribbons and the mean shear within the PIL.  Using a linear function we find the $(13\pm3)x+(6\pm5)$ relationship between the two and a Spearman correlation coefficient of $0.7$. We find a similar relationship between the AR net current and the PIL shear, with a slightly weaker correlation coefficient. What does this relationship imply? It implies that the net current is a manifestation of the magnetic shear at the PIL.


\subsection{FlareMagDB Statistics: Confined vs. Eruptive Events}\label{conf}
We analyze magnetic field properties of $7$ {\it confined} events, that did not cause CMEs (marked with $\bigtriangleup$s in Figure~\ref{4by4} and Figure~\ref{2by2}), and compare them with $33$ {\it eruptive} events, that caused CMEs (marked with $\bullet$s). $7$ confined events originated in $5$ ARs, including $3$ X-class flares, $X3.1$, $X1.6$ and $X1.0$, originating in the AR 12192 \citep{Sun2015}.  

\begin{figure*}[tb!]
  \centering
 \resizebox{1.0\hsize}{!}{\includegraphics[angle=0,trim=3cm 0cm 3cm 0cm,clip,width=\textwidth]{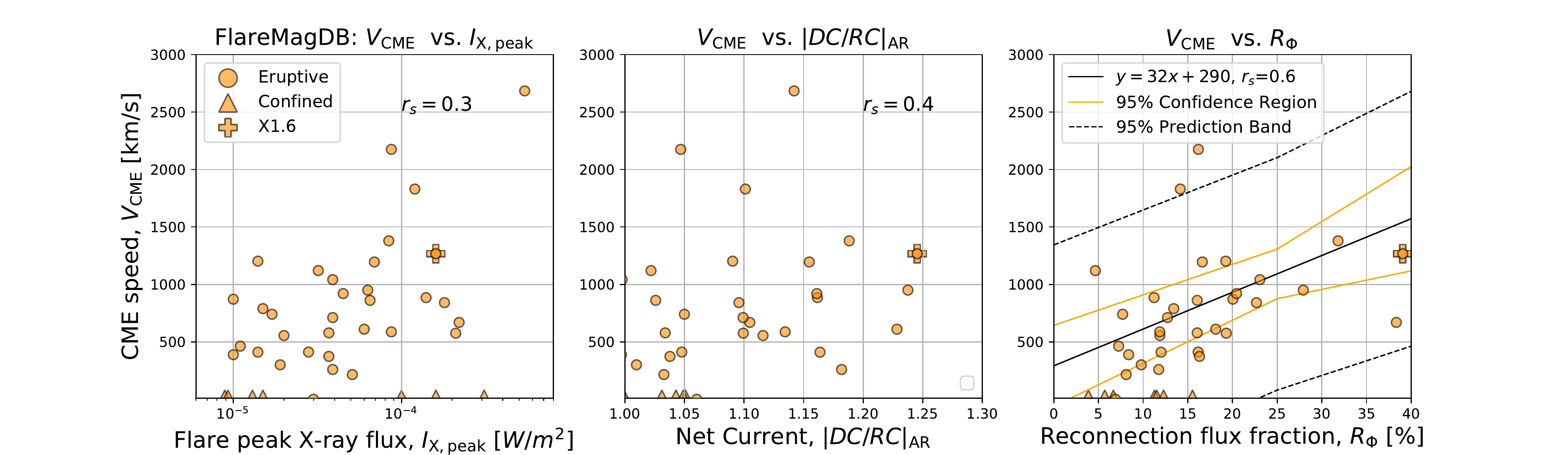}}
  \caption{{\it FlareMagDB results}: CME speeds  vs. highest-correlated flare magnetic properties, peak X-ray flux, AR net current and the reconnection flux fraction, for 33 eruptive ($\bullet$) and 7 confined ($\bigtriangleup$) flares of FlareMagDB database. See \S\ref{conf}.}
  \label{3by1}
\end{figure*}

Comparing confined vs. eruptive events, we find several differences in their magnetic field properties.
First, we find that for flares within a certain peak-X-ray-class range, confined events have larger PIL magnetic fluxes than eruptive events (Figure~\ref{4by4}a).  
Second, we find that confined events have smaller PIL-shear and ribbon net current than eruptive events (cf. $\bigtriangleup$ vs. $\bullet$).
Third, we find that the reconnection flux fraction, $R_\mathrm{\Phi}$, tends to be smaller for confined events than for eruptive events  (Figure~\ref{4by4}b). If we look at eruptive events alone, we find that their reconnection flux fractions strongly correlate with the CME speeds, $r_s(r_\Phi,v_\mathrm{CME})=0.6$. In fact, among all \verb+FlareMagDB+ variables, it is the reconnection flux fraction that has the strongest correlation with the CME speed. In Figure~\ref{3by1} we plot the CME speed vs. the \verb+FlareMagDB+ variables that have the highest correlation coefficients with the CME speed, peak X-ray flux, AR net current and the reconnection flux fraction.  
To summarize, we find that while PIL shear and the net current have weak to moderate correlation with the CME speed, the reconnection flux fraction has a strong correlation with the CME speed, i.e. the properties of the field of the flux rope play smaller role in the CME speed that the properties of flux rope relative the overlying field.

We note that our analysis of field properties within confined and eruptive events is based on $33$ eruptive and only $7$ confined events. Analysis of a larger data sample with equal numbers of confined and eruptive events would be necessary to draw more certain conclusions.

%
\section{\textsl{Results: 3D MHD ARMS Simulations}}\label{prelim-mhd}
%
To provide theoretical ground for our observational findings, here we present the same observational analysis on the results of a 3D MHD simulation of an idealized magnetic breakout CME \citep{Antiochos1999,Lynch2008} that forms a large-scale post-eruption arcade below an energetic flux rope (FR) eruption. 
We note, that while the simulation's AR magnetic flux is similar to a flux of the medium active region, the mean magnetic field strengths (and AR size) are about $100$ times smaller (larger) in the simulations than observed. Using the scaling law between the reconnection flux and the peak X-ray flux, we conclude that the simulation's reconnection flux corresponds to an M1.0-class flare.

\subsection{Model Details and Initial Conditions}

The 3D MHD simulation was performed with the Adaptively Refined MHD Solver \citep[ARMS;][]{DeVore2008}. The simulation is a  left-handed \citet{Lynch2009} configuration, energized with a pair of idealized shearing flows parallel to the active region PIL.

The magnetic field at $t=0$~hr is initiated via a sum of point dipoles,
\begin{equation}
\mathbf{B}(\mathbf{r},0) = \sum_i M_i \left( \frac{R_i}{{r'_i}} \right)^{3} \left[ \, 3\mathbf{n}_i \left( \mathbf{n}_i \cdot \mathbf{m}_i \right) - \mathbf{m}_i \, \right] \; ,
\end{equation}
where $r'_i = | \mathbf{r}-\mathbf{r}_{i,0} |$, $\mathbf{n}_i = \mathbf{r'}_{i}/{r'_i}$, and each dipole has a moment magnitude $M_i$, location $\mathbf{r}_{i,0}$, scaling factor $R_i$, and is pointing in the $\mathbf{m}_i$ direction.
As in \citet{Lynch2009}, $i=[1, \, 6]$, and the parameters used here are given in Table~\ref{tab:arms}.

The base pressure and temperature are given by \textcolor{black}{$p_0 = 0.025$~dyn~cm$^{-2}$ and $T_0 = 1.9433 \times 10^6$~K}, respectively, while the initial solar atmosphere is in hydrostatic equilibrium with $p(r) = p_0 \left( r/R_\odot \right)^{-\mu}$, $T(r) = T_0 \left( r/R_\odot \right)^{-1}$, $p = 2 ( \rho / m_p ) k_B T$, and \textcolor{black}{$\mu = R_\odot/H_\odot = 12.0$} is the normalized (inverse) scale height. This yields a base mass density of \textcolor{black}{$\rho_0 = 7.80 \times 10^{-17}$~g~cm$^{-3}$ corresponding to a base number density of $n_0 = 4.6634 \times 10^7$~cm$^{-3}$}.
 
The computation grid is block decomposed with an initial resolution of \textcolor{black}{$5 \times 5 \times 5$ blocks covering the full domain of $r \in [1R_\odot, \, 20R_\odot]$, $\theta \in [ 0.0625\pi, 0.9375\pi]$ (latitude of $\pm78.75^{\circ}$) and $\phi \in [-0.5\pi,+0.5\pi]$}.
Each block contains $8^3$ grid cells and we have employed 4 additional levels of static grid refinement. The highest-refined region covers  \textcolor{black}{$r \in [1R_\odot, \, 6R_\odot]$, $\theta \in [ 0.32\pi, 0.55\pi]$ (latitude of $[ -9.0^{\circ}, +32.4^{\circ} ]$) and $\phi \in [ -0.18\pi, +0.18\pi ]$ with an effective resolution of $320^3$. Thus, the highest-refined grid cells have a $0.49^{\circ} \times 0.56^{\circ}$ angular resolution in $(\theta,\phi)$ and $\Delta r = 0.0188R_\odot$ at the lower boundary}.
 
The energizing boundary flows impart a significant shear component ($B_\phi$) to the active region as each polarity concentration is moved further apart. \textcolor{black}{The form of the \citet{Lynch2009} shearing flows are
\begin{equation}
\mathbf{v}_h^{(\pm)} = \pm v_0 f^{(\pm)}(\theta, \phi) \left( \frac{1}{2} - \frac{1}{2} \cos{\left[\frac{2\pi t}{(10^4~{\rm s)}} \right]} \right) \, \hat{\phi} \; ,
\end{equation}
where $(\pm)$ represents the flows applied to the positive and negative polarities, respectively, for the duration $t \in [0,2.778]$~hr.}
The function $f^{(\pm)}(\theta, \phi)$ smoothly ramps the flow profile to zero outside of specified $(\theta, \, \phi)$ ranges on either side of the PIL, described in \citet{Lynch2009}. The magnitude of the shearing flows \textcolor{black}{$v_0 = 65$~km~s$^{-1}$} is significantly faster than observed photospheric velocities but below the sound speed
\textcolor{black}{($c_0 \sim 180$~km~s$^{-1}$)}
and extremely sub-Alfv\'{e}nic  
\textcolor{black}{($v_A \gtrsim 2000$~km~s$^{-1}$)}
in the vicinity of the AR.

\begin{deluxetable}{rrrrr}
\tablecaption{ARMS initial magnetic field point dipole parameters\label{tab:arms}}
\tablewidth{0pt}
\tablehead{
\colhead{$i$} & \colhead{$M_i$} & \colhead{$R_i$} & \colhead{$\mathbf{r}_{i,0}$} & \colhead{$\mathbf{m}_{i}$} \\[-0.05in]
\colhead{} & \colhead{[G]} & \colhead{$[R_{\odot}]$} & \colhead{$(\, r/R_\odot,\, \theta,\,\phi \,)$} & \colhead{$(\, \hat{r},\, \hat{\mathbf{\theta}},\, \hat{\mathbf{\phi}} \, )$}
}
\startdata
1  &   \textcolor{black}{1.00}  &  1.0  &  ( 0, 0, 0 )  &  ( 0, 1, 0 )  \\
2  &   \textcolor{black}{0.08}  &  1.0  &  ( 0.80, 0.45$\pi$, $-0.025\pi$ )  &  ( 0, 1, 0 )  \\
3  &   \textcolor{black}{0.08}  &  1.0  &  ( 0.80, 0.45$\pi$, $+0.025\pi$ )  &  ( 0, 1, 0 )  \\
4  &   \textcolor{black}{0.10}  &  0.33  &  ( 0.95, 0.45$\pi$, 0 )  &  ( 0, 1, 0 )  \\
5  &   \textcolor{black}{0.10}  &  0.33  &  ( 0.95, 0.45$\pi$, $-0.010\pi$ )  &  ( 0, 1, 0 )  \\
6  &   \textcolor{black}{0.10}  &  0.33  &  ( 0.95, 0.45$\pi$, $+0.010\pi$ )  &  ( 0, 1, 0 ) \\
\enddata
\end{deluxetable}
 
\subsection{Eruption Details} \label{arms_det}
The magnetic breakout model for CME initiation relies on the positive feedback process associated with reconnection at a current sheet \emph{above} the source region's expanding sheared core \citep{Antiochos1999,Lynch2008}. The arcade expansion accelerates as the restraining over-lying flux is transferred into the adjacent arcades of the AR multipolar flux system. 
Reconnection at the breakout current sheet drives the initial stages of the eruption and leads to the transition from slow quasi-ideal evolution to a driven, runaway configuration that forms a vertical current sheet \emph{below} the expanding sheared field core.
 This current sheet thins and elongates with continued expansion and, with the onset of the fast CSHKP eruptive flare reconnection, ushers in the explosive acceleration phase of the CME eruption \citep[e.g.][]{DeVore2008, Karpen2012, Lynch2013, Dahlin2019, Wyper2021}.

\begin{figure*}
    \includegraphics[width=1.0\linewidth]{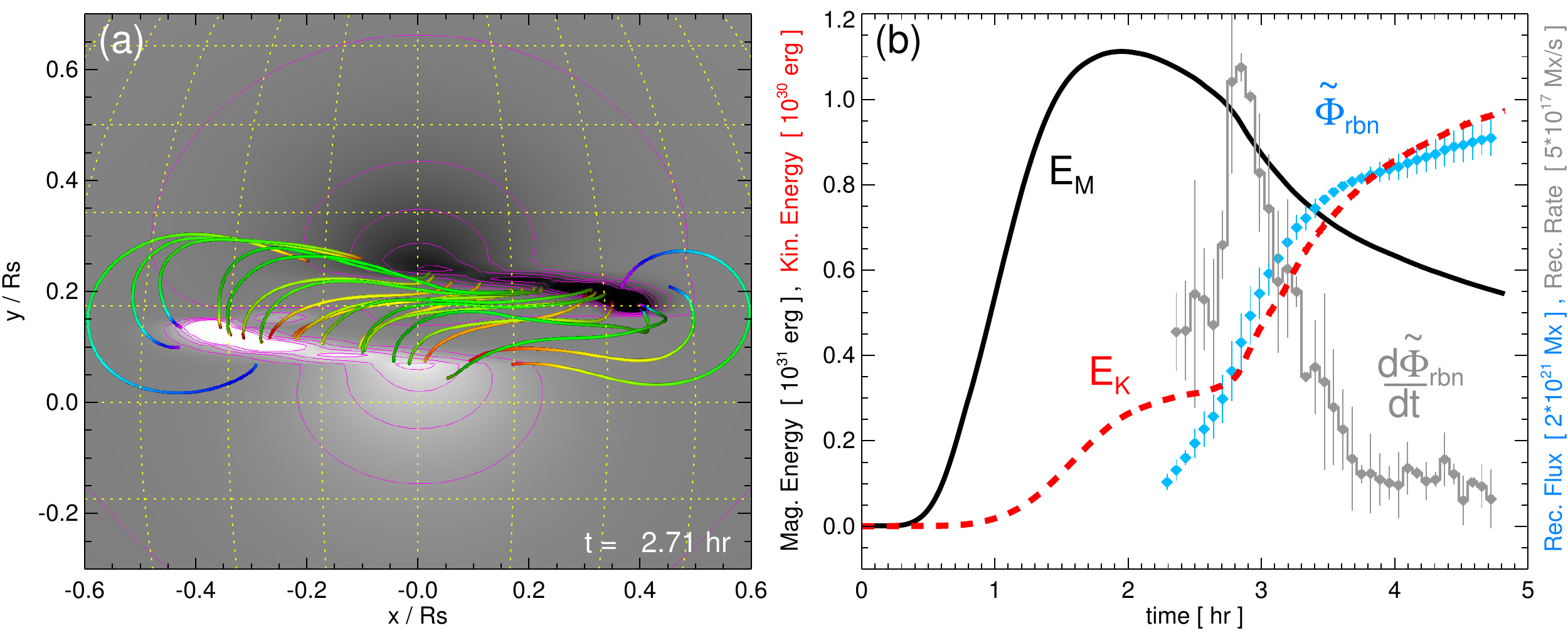}
    \caption{ARMS simulation results during the CME eruption: {\it Panel a}: radial magnetic field on the $r=R_\odot$ lower boundary with post-eruption arcade field lines at $t=2.71$~hr. {\it Panel b}: total magnetic, $E_M(t)-E_M(0)$, and kinetic, $E_K(t)$, energies along with total unsigned reconnection flux ,$\Phi_{\rm rbn}$, and reconnection rate, $d\Phi_{\rm rbn}/dt$. See \S\ref{arms_det} for details.}
    \label{fig_arms1}
\end{figure*}

Figure~\ref{fig_arms1} shows a snapshot of the simulation results during the CME eruption. Panel~\ref{fig_arms1}a shows the field lines of the post-eruption flare arcade in the very beginning of the eruption ($t=2.71$~hr), and panel~\ref{fig_arms1}b shows the temporal evolution of total magnetic energy, kinetic energy, flare reconnection flux, and reconnection rate. To focus on changes of the magnetic energy during the eruption we have set the magnetic energy at $t=0$ to zero: $E_M(t=0)=0$. The amount of the total released magnetic energy, i.e. the $E_M$ decrease from $2$ to $5$hr, is $\Delta E_M \sim 6 \times 10^{30}$~erg, which roughly corresponds to the change in the free magnetic energy.  
The kinetic energy exhibits two stages of evolution. During the initial stage from $1$~hr to $2.6$~hr the kinetic energy $E_K$ increases slowly corresponding to a slow quasi-ideal rise of the sheared core before the onset of flare reconnection. This expansion is initially caused by the force imbalance (magnetic pressure gradient) introduced with the sheared field component. 
During the second runaway stage, from $2.6$~hr to $5$~hr, $E_K$ increases much more rapidly corresponding to the start of the  breakout reconnection.  Once the flare reconnection starts, it becomes the dominant process driving the entire subsequent evolution of the CME eruption \citep{Karpen2012}. By the end of the flare at $t=5$~hr, the kinetic energy reaches a maximum of $E_K \sim 10^{30}$~erg and the cumulative unsigned reconnection flux reached $\Phi_{\rm rbn} \sim 1.8\times10^{21}$~Mx. The maximum reconnection rate is  $d\Phi_{\rm rbn}/dt \sim 5.5 \times 10^{17}$~Mx~s$^{-1}$ at \textcolor{black}{$t = 2.85$~hr}. The total fraction of the active region flux that participated in the flare is $43\%$ (Table~\ref{summary_table}).
 
\subsection{Magnetic Field and Electric Current Properties in the AR, PIL, and Flare Ribbon Regions}\label{arms_obs}

To apply the observational analysis framework to the simulation results, we first project the spherical magnetic field components into a Cartesian coordinate system where $( x,\, y )$ represent the plane-of-the-sky (horizontal components) and $z$ comes out of the page towards the observer (vertical component).
%
Next, we interpolate the simulation values onto a uniform $600 \times 500$ array corresponding to the area shown in Figure~\ref{fig_arms1}a, i.e. $x \in [ -0.6R_\odot, \, +0.6R_\odot ]$, $y \in [-0.25R_\odot,\, +0.75R_\odot]$.
This yields a pixel area $dA_{ij} = \Delta x \Delta y = 1.96 \times 10^{16}$~cm$^2$.
From this data set, we calculate the magnetic shear $\alpha(x,y)$, vertical current density $J_z(x,y)$, and their respective averages over the regions of interest defined in Section~2.2.

\begin{figure*}
    \includegraphics[width=1.0\linewidth]{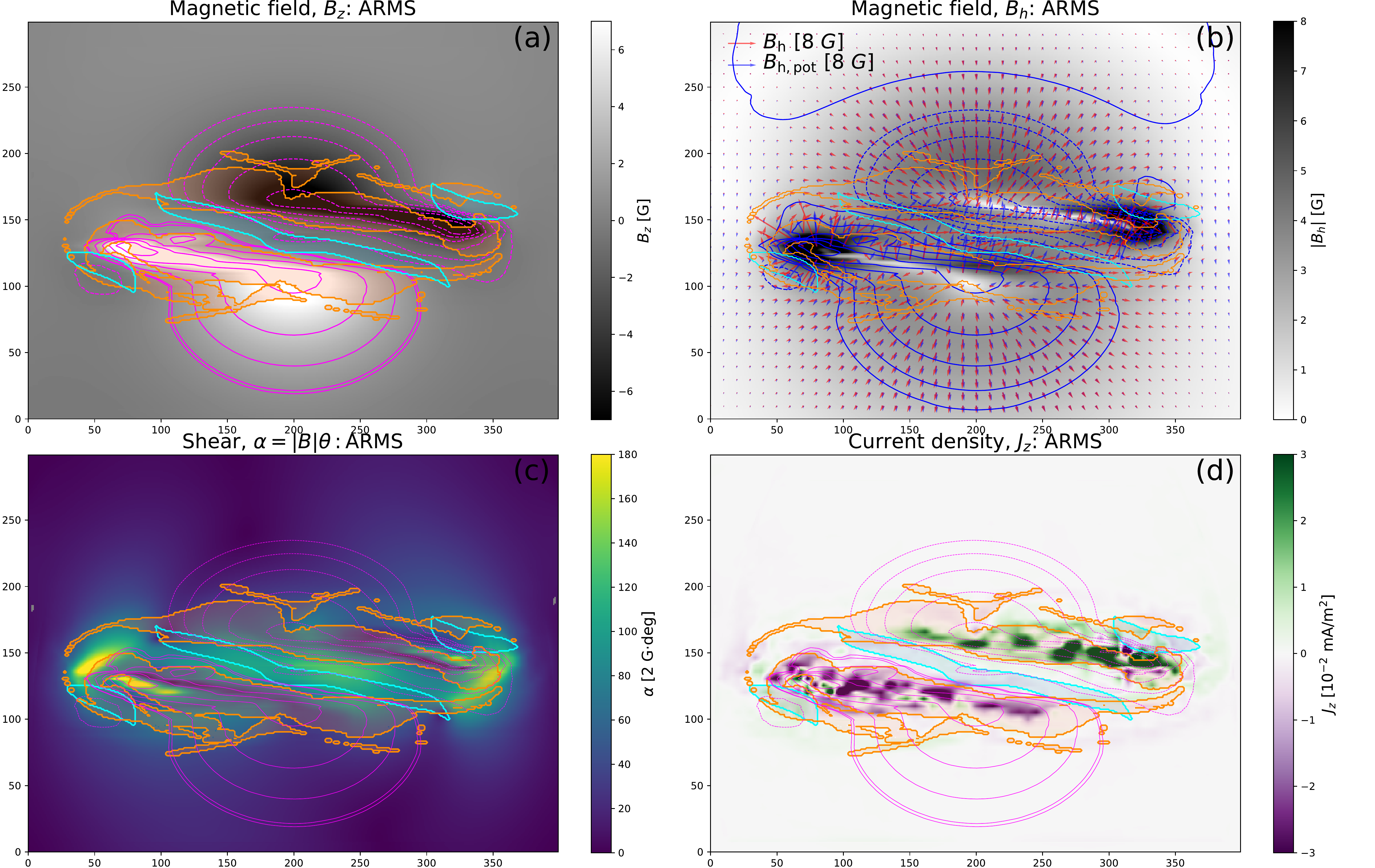} 
    \caption{ARMS MHD quantities on the $r=R_\odot$ lower boundary analogous to those presented for an observed X1.6 flare in Figure~\ref{db_example}. {\it Panel a}: vertical field $B_z$.  {\it Panel b}: magnitude of the horizontal magnetic field and vector directions of the total and potential components (arrows).  {\it Panel c}: distribution of magnetic shear $\alpha(x,y)$. {\it Panel d}: vertical current density $J_z(x,y)$. In each panel the total flare ribbon area during the eruption is denoted as the orange-shaded region while the cyan-shaded region indicated the AR PIL. See \S\ref{arms_obs} for details. }
    \label{fig_arms2}
\end{figure*}

Figure~\ref{fig_arms2} shows the ARMS simulation version of Figure~\ref{db_example} at 
{$t= 2.57$~hr}, {\it right before the onset} of the fast eruptive-flare reconnection. Figures~\ref{fig_arms2}a, b show the vertical and horizontal field components with the cumulative, over the flare duration, ribbon and PIL areas denoted as the orange and cyan shading, respectively. Here the ribbon areas are calculated from a field line length proxy, as in \citet{Lynch2019,Lynch2021}. Figure~\ref{fig_arms2}c shows the distribution of shear and Figure~\ref{fig_arms2}d shows the distribution of the vertical current density.
 
Table~\ref{summary_table} shows a comparison of observed values from \verb+FlareMagDB+ with ARMS.
The mean magnetic flux in each region is $\Phi(AR,rbn, PIL) = [4.1,1.8,0.03]\times10^{21}$ Mx.
The mean shear in each region is $\bar{\alpha}(AR,rbn,PIL)=[63,129,169]\times (G\cdot deg)$.
The total unsigned vertical current in each region is $I _{\rm z,u}(AR,rbn,PIL) = [2.7,2.1,0.03]\times10^{11}$ A.
The net current in each region is $|DC/RC|(AR,rbn,PIL) = (4.9,10.5,3.1)$.
Comparing these values with observations, ARMS magnetic fluxes correspond to a medium-size flare in a medium-size active region. For example, an M1.0-class flare in a sigmoidal AR ($i=0$) of \verb+FlareMagDB+ has very similar AR, ribbon and PIL magnetic fluxes. 
The size of the ARMS AR is much larger than any observed case, therefore the magnetic fields are much weaker. Since the magnetic fields are much weaker, ARMS magnetic shears (which is a product of the magnetic field magnitude and the shear angle, $\theta$) and total currents are much smaller than the observed ones. We also find that ARMS AR has very weak return currents, resulting in large net currents, which vary from $5$ for a single polarity to $10.5$ within a flare ribbon. For comparison, the observed ARs have smaller net currents ranging from $1$ to $3$, i.e. the return currents in the observations are stronger. Note that before the eruption when we start shearing the arcade, there is direct current close to PIL and a clear return-current shell separating the sheared from un-sheared components along the quasi-separatrix layer. During the eruption the return current weakens as a result of peeling away of the outer layers of the expanding flux system via breakout reconnection. 

To summarize, from ARMS analysis we find the following: 
(1) Similar to observations, simulations' mean magnetic shear increases gradually from AR to ribbon and PIL areas. Simulations' shear values are consistent with the observed ones, given $100$-times scaling relation between mean magnetic fields.
(2) Similar to observations, the vertical electric currents within individual polarities are highly non-neutralized, with the largest net currents located within areas involved in the eruption. Simulation's net currents are $\approx5$ times stronger  than observed values. (3) We find a close spatial relationship between the simulation's magnetic shear at PIL and the net current, confirming the new net current-shear scaling law we find from the observations.

\section{Discussion}\label{disc}

We used HMI/SDO photospheric vector magnetic field observations in $40$ flares to assemble a  \verb+FlareMagDB+ database with properties of magnetic fields within AR, flare ribbon and PIL areas. We analyzed pre-flare {\it observed} magnetic flux, reconnection fraction, magnetic shear, total electric current and net current for $40$ events and compared these with flare X-ray peak flux (Figures~\ref{4by4} and \ref{2by2}) and CME speed (Figures~\ref{3by1}). In Table~\ref{summary_table} we summarized variables' typical range and compared these with synthetic variables from an MHD ARMS eruption simulation.  In the Correlation Matrix~in Figure~\ref{heatmap}  we investigated dependencies between the variables. In this section we describe how our results complement existing understanding of flare magnetism.

{\it Ribbon vs. PIL Morphology:} We compare spatial distribution of ribbons and PILs and find that many flares, including M5.0 and above, have ribbons lying away from the PIL. What does it imply? First, that for many events the morphology of PILs has almost {\it no bearing} on the morphology of flare ribbons. Second, that in these events low-lying magnetic fields that are rooted in the vicinity of the PIL area do not participate in the flare. How do these ideas fit within the contention that large near-PIL fluxes indicate that a large flare is more likely to occur \citep{Schrijver2007}? We explain this with the fact that the majority of events in our list,  $33$ out of $40$, have sigmoidal coronal structure \citep{Canfield2007,Savcheva2014}.
Sigmoid is a twisted FR embedded in a surrounding field. \citet{Savcheva2012a,Savcheva2012b} analyzed the time-evolution of a sigmoid eruption and constructed a scenario for the onset of the CME in sigmoidal regions: the pre-existing FR can develop into a CME as a consequence of reconnection at a Hyperbolic Flux Tube (HFT) under the FR, which increases the magnetic flux in the rope and at the same time weakens the overlying magnetic arcade holding it down. Thus, tether-cutting reconnection is needed in the slow rise phase of the eruption, and the torus instability is necessary to allow the CME eruption.
On the other hand, presence of ribbons close to the PIL indicates of either the presence of a low-lying pre-existing flux rope or a creation of a FR during the flare as a result of the reconnection of the low-lying sheared magnetic fields \citep{Kazachenko2012}. A more detailed study of flare ribbon vs. PIL position and PIL properties relative to coronal structure would be helpful to advance our understanding of flux rope formation.

{\it Magnetic flux:}
From the analysis of magnetic fluxes we find that ribbon and PIL fluxes have strong correlation with the AR flux, implying that large PIL and ribbon fluxes tend to occur in active regions with large magnetic fluxes.  We also find that magnetic fluxes have moderate to strong correlation with other magnetic field products like mean magnetic shear and the total current. Specifically the correlation between the PIL flux and the AR shear is very strong ($r_s=0.8$, see Correlation Matrix~\ref{heatmap}). One possible scenario to explain this relationship --  as the active region evolves and gets pushed by photospheric flows, the AR field becomes more sheared and magnetic flux piles up at PIL due to, for example, collisional shearing \citep{Chintzoglou2019}.

We find strong correlation between the AR, ribbon and PIL magnetic fluxes and the peak X-ray flux ($r_s>0.6$). While strong, correlation between the PIL flux and the X-ray peak flux is weaker than the correlation between the PIL flux and the AR flux ($r_s=0.6$ vs. $r_s=0.8$). In other words,  while correlation between PIL flux and flare size is in line with \citet{Schrijver2007} hypothesis that flare-productive ARs tend to have stronger PILs (see also references in \citealt{Georgoulis2019}), this correlation could reflect the fact that larger ARs have stronger PILs and hence host larger flares.  A study of small vs. large active regions with the same amount of PIL flux might be useful to find out what plays the major role in the AR flare production, the AR flux or the PIL flux. We also find a very strong correlation between the reconnection flux and the peak X-ray flux ($r_s=0.9$), in agreement with earlier results for larger datasets \citep{Kazachenko2017,Toriumi2017,Tschernitz2018}. 
The power law between the two quantities agrees with the \citet{Warren2004} scaling from hydrodynamic simulations of impulsively heated flare loops, proving that the soft X-ray radiation energy released during the flare comes from the free magnetic energy released during reconnection \citep{Kazachenko2017}.

{\it Magnetic flux fractions:} We find that fractions have a moderate correlation with the peak X-ray flux ($r_s=0.5$) that is weaker than correlation between the ribbon flux and the peak X-ray flux. This result is in agreement with \citet{Kazachenko2017} ($r_s=0.5$) and \citet{Toriumi2017}. We explain this relationship as a consequence of the way how we calculate the fractions ($R_\Phi=\frac{\Phi_\mathrm{ribbon}}{\Phi_\mathrm{AR}}\times100\%$) and a very strong correlation between the ribbon flux and the peak X-ray flux.

{\it Mean magnetic shear:}
The amount of magnetic shear at PIL is known to be related to the flare productivity of an active region. Here, as noted above, we find that the amount of magnetic shear within the active region is most strongly correlated with the flux at PIL which in turn is strongly correlated with AR magnetic flux. The larger is the active region, the larger is the amount of shear it contains and more flux it has at PIL. Is the amount of shear related to flare peak X-ray flux? From our analysis, we find that it is not the case -- mean magnetic shear has weak to moderate correlation with the peak X-ray flux that is weaker than the correlation between magnetic flux and the peak X-ray flux. For example, while PIL flux correlates with the flare size ($r_s=0.6$), increased PIL shear does not correlate with AR flare size ($r_s=0.2$). We also find that this correlation decreases from ARs to ribbons and PILs. As we go further from the PIL, we get less sheared  magnetic fields, consistent with the standard 3D flare model \citep{Janvier2014,Savcheva2016}.
Analysis of  many large flares with large amount of PIL magnetic flux and small amount of shear at the PIL, such as for example an M8.7 flare ($i=8$, $\Phi_\mathrm{PIL}=21.2$) or an M6.5 ($i=15$, $\Phi_\mathrm{PIL}=11.7$) would be interesting.

In several past investigations and in ongoing work, \citet{Savcheva2015,Savcheva2016,Janvier2016,Karna2021} have studied several of the active regions shown in Table~\ref{events}, precisely, 2010-08-07 ($i=0$), 2011-02-15 ($i=1$), 2011-09-06 ($i=3$), 2012-03-07  ($i=9$), 2012-03-09 ($i=10$), 2012-07-12 ($i=13$), 2013-04-11 ($i=15$) and 2014-10-22-25 ($i=31,32,33$). Specifically, \citet{Savcheva2015,Savcheva2016} studied an M1.0 flare on 2010-08-07 (i=0) using a magnetofrictional (MF) evolution. They found that first reconnection happens under the flux rope turning two J-shaped field lines from the sheared arcade into an S-shaped field line that builds magnetic flux in the flux rope and produces flare field lines as a by-product. These flare field lines, that match the observed flare loops, initially lie close to the PIL and are highly sheared, carrying remnant shear of the flux rope shear, and sequentially, as the flux rope goes up in height, less and less sheared loops form that are rooted farther and farther away from the PIL  (see cartoon in Fig 1. from \citealt{Savcheva2016}). This scenario, using MF simulation, supports data-constrained MHD simulation by \citet{Kliem2013} and also the observed large to small shear transition from PIL to ribbon areas that we find here.
{\it  Spatial structure of the vertical current density:}
What is the spatial structure of vertical current density and does it have any relationship with the ability of the AR to host a CME vs. a confined event? Recent works suggest that CME-eruptive ARs exhibit defined filamentary structures or ribbons in the current density maps while CME-quiet ARs do not (see e.g. Figure 3 of \citealt{Avallone2020}).  Some observed case studies and simulations suggested that a hook-shaped pattern in these filaments could be a signature of the pre-existing  flux rope \citep{Schmieder2018,Janvier2016,Barczynski2020}. Do we see direct current ribbon pairs aligned with PIL before an AR hosts a CME (similar to the $q=0.8$ case in \citealt{Sun2021})? Do these current ribbons have hook shapes? To answer this question we examine \verb+FlareMagDB+ current density maps in Figure~\ref{jz_examples}.  First, we notice that few of these events have a clear hook shape. I.e. while hooks might be a signature of a pre-existing flux rope, hook presence is a rare phenomenon. Second, we find that while some eruptive events are associated with distinct current ribbon pairs close to PIL, others do not have these current ribbon pairs, instead consisting of salt-an-pepper patterns of positive and negative current density.  For example,  AR 11158 ($i=1$, eruptive X2.2 flare) and AR 12241 ($i=36$, eruptive M6.9 flare) have clear current ribbon pairs, while AR 11944 ($i=20$, eruptive X1.2 flare) does not have current ribbons. Does it mean that eruptive events that do not have current ribbons are all in-situ formed flux ropes, i.e. flux ropes that were primarily formed during reconnection?  Finally, we find that all seven confined events in our list do not have current ribbons.  This result is consistent with earlier works and the scenario that lack of current ribbons might be related to absence of the pre-existing flux rope and as a result smaller likelihood of the flare to erupt. We conclude that a more detailed morphological study of $J_z$ patterns in ARs hosting eruptive and confined events with analysis of pre-flare coronal images to track FR existence is needed to clarify the relationship between $J_z$ morphology and the AR ability to erupt.  

We also calculate the total and mean vertical current within AR, ribbon and PIL areas. We find that the total vertical current increases as we go from PIL to AR areas, reflecting larger integration areas. On the other hand, the mean current decreases from PIL to AR areas, since within PIL and ribbon areas magnetic fields are much more non-potential than in the AR as a whole.

{\it  Net current within AR, ribbons and PILs:} 
Understanding how currents are neutralized is important for numerical models of solar eruptions (see introduction in \citet{Torok2014} paper). If we look at the AR as a whole, we find that $|DC/RC|\approx1$, i.e the current is neutralized, in agreement with previous studies \citep{Wheatland2000,Venkatakrishnan2009,Georgulis2012}. On the other hand, individual polarities contain a net current. This net current varies from $1$ to $3$  and is largest in the central part of the active region, around PIL and ribbon areas, decreasing gradually within the polarity as a whole (an AR net current). In the simulations, net currents are larger but exhibit similar growth from the periphery to the center of the AR. For example, using ARMS simulations, here we find net currents reaching up to $5$ within individual AR polarities and $10$ within the ribbon areas. For a simulation of an emerging active region, \citet{Torok2014} found net currents reaching up to $5$, increasing as an AR emerges, consistent with pre-flare ARMS values.
\citet{Titov2008} and \citet{Aulanier2010} in their MHD simulations also found weak return currents with net currents around $3$, similar to observed values we find here with \verb+FlareMagDB+.

{\it What creates these net currents?} Net currents could emerge bodily into the corona (''emergence'', e.g. \citealt{Leka1996,Longcope2000}) or be produced by stressing of the coronal magnetic field by sub-photospheric and photospheric shearing flows (``shear'', e.g. \citealt{McClymont1989,Torok2003,Aulanier2010}).
From our statistical analysis we find that the net current within each AR polarity (or ribbons) is proportional to magnetic shear at PIL, with most of the direct current lying in the PIL vicinity. In other words, net current is a manifestation of the magnetic shear at the PIL. This conclusion is in agreement with recent findings using simulations \citep{Dalmasse2015,Torok2014} and one-case {\it morphological} studies using observations \citep{Ravindra2011, Georgoulis2012,Vemareddy2015,Liu2012}.  \citet{Torok2014,Dalmasse2015}  found that net currents are co-temporal with surge of shear at PIL and can be formed due to flux emergence, twisting and shearing motions and any mechanism that can generate magnetic shear along a PIL. Using high-resolution vector magnetograms from the Spectropolarimeter/Hinode instrument of an emerging AR over several days, \citet{Ravindra2011} noticed that  net current within individual polarities could be related to shear at PIL.  In a different study, \citet{Venkatakrishnan2009} found that non-sheared PILs have zero net current. In this paper we perform a first  {\it quantitative} analysis of a fairly large observations dataset and find that the net current is proportional to shear at PIL. This finding is expected from the integral form of the Amp\`ere's law, $\oint {\bf B}\cdot dl = (4 \pi / c) I$. While along the PIL $\oint {\bf B}\cdot dl>0$,  along the curved segment away from the PIL,  $\oint {\bf B}\cdot dl\approx0$.
The presence of net current in our sample of ARs supports previous studies that state that return currents get trapped under the photosphere during flux emergence, leading to the presence of a net current causing increased shear at PIL. Our results also confirm that coronal flux rope models that neglect return currents are a valid representation of pre-eruption configurations on the Sun.


{\it Confined vs. eruptive flares:}
Here we find that for a given X-ray flux, confined flares have smaller reconnection flux fractions, $R_\mathrm{\Phi}$, (cf.  $\bigtriangleup$ vs. $\bullet$ in Fig.~\ref{4by4}b). 
 If we go one step further and look not just at the flare eruptivity, but at the CME speeds, we find that among all magnetic variables of \verb+FlareMagDB+, reconnection flux fraction is the only variable that has a strong correlation with the CME speed ( $r_s=0.6$, see Correlation Matrix~\ref{heatmap}). Earlier, some indications of this relationship were suggested by \citet{Tschernitz2018} and \citet{Toriumi2017,Li2020}.  \citealt{Tschernitz2018} analyzed $51$ flares ($32$ confined, $19$ eruptive) ranging from GOES class B3 to X17. They found that confined flares of a certain GOES class have smaller ribbon areas, but larger mean magnetic fields. This result implies that confined flares occur closer to the flux-weighted center of ARs, where fields that could be swept by the flare ribbons could be stronger. In a different study, \citet{Toriumi2017,Li2020} compared $R_\mathrm{\Phi}$ in confined vs eruptive flares, finding that eruptive events have on average larger flux fractions. 
 Here, instead of a binary comparison between confined and eruptive events properties,  we for the first time analyzed the CME speed. In Figure~\ref{3by1} we show the scatter plot of the CME speed vs. the reconnection fraction with the functional fit of
 $y=ax+b$ describing the relationship, where $a=32\pm9$ and $b=290\pm180$ and $R^2=0.3$. If correct, the new relationship between the CME speed and the reconnection fraction could in principle be used to predict the CME speed once the solar flare occurs.

How could we explain the correlation between the CME speed and the reconnection flux fraction? We suggest the following scenario. Once the flare starts, reconnection starts feeding magnetic flux into a pre-existing flux rope or a sheared arcade.  As this new flux rope forms, overlying fields within an AR provide a retrieving force to keep this flux rope stable. As a result the larger is the reconnection flux fraction, the weaker is the restraining arcade above and the faster the new flux rope erupts reaching torus instability at a lower height \citep{Chen2019,Kliem2021}.  The torus instability occurs when the overlaying potential arcade decays with a critical index, $n_\mathrm{crit}=-\frac{d ln B}{d ln z} \approx 1.1 - 1.5$, depending on the flux rope geometry and shear in the overlaying arcade \citet{Kliem2006}. If the tension of the arcade above is not sufficient, the flux rope starts expanding, thinning the reconnection current sheet further and causing more reconnection to take place. This scenario accounts for the slow rise of filaments based on tether-cutting-like reconnection \citep{Moore2001} and the fast eruption after the torus instability kicks in. It is also in line with a recent observation analysis by \citet{Li2020,Li2021} who analyzed 322 and 719 flares, respectively, and found that the flare-CME association rate decreases with the increasing magnetic flux of the AR that produces the flare, implying that large magnetic flux tends to confine eruptions.

Physically reconnection flux fraction represents the ratio between the flux gained by the non-potential flux rope to the total more potential flux of the active region. \citet{Lin2021} generalized this idea calling it ``relative non-potentiality'' of a magnetic flux rope. Consistent with the scenario above, a larger relative non-potentiality indicates a higher probability for a flux rope to erupt. \S1 of \citet{Lin2021} contains a nice overview of a variety of schemes to quantitatively evaluate the relative non-potentiality.

We find that all of the confined events in our dataset have reconnection flux fraction, $R_\Phi<15\%$, below the instability criteria, $R_\mathrm{\Phi,inst}\approx30\%$, determined by the use of the flux rope insertion method \citep{Savcheva2009,Savcheva2012,Su2011}. (For comparison, the typical $20^{th}$ to $80^{th}$ percentile for the reconnection flux fraction in \verb+FlareMagDB+ is $R_\Phi[P_{20},P_{80}]=[8.2,20.1]\%$ and the ARMS value is $R_\Phi\approx43\%$). On the other hand, as seen from the CME speed vs. reconnection flux fraction plot in Figure~\ref{3by1}, there are many  events that are eruptive and yet have the reconnection flux fraction below the instability criterion, $R_\mathrm{\Phi,inst}\approx30\%$. A systematic analysis of these outlier events would be interesting to understand why these events are eruptive.

Compared to eruptive events, we find that confined events tend to have stronger PIL fluxes. We speculate that for a constant amount of flux in the flux rope, stronger vertical PIL flux implies weaker horizontal flux and as a result smaller magnetic energy within a flux rope.

We also compare PIL shear (and a related ribbon net current) for confined vs. eruptive events and find that confined events have smaller PIL-shear and ribbon net current. This result statistically confirms earlier studies of  \citet{Liu2017} (4 events), \citet{Avallone2020} (30 events) and \citet{Kontogiannis2019} (32 events) who found that flare-active/CME-active ARs are associated with larger net currents than flare-quiet/CME-quiet ARs.  For our limited dataset, however, this relationship is not very strong with large number of outliers, i.e. there are many eruptive events with small net current, as has been previously noted by \citet{Avallone2020}. Case studies of these events might be useful to understand the relationship between eruptivity and the net current. From the physics standpoint, we explain the lack of shear in confined events by a smaller non-potential energy and hence a smaller likelihood to erupt. 

We note that all our conclusions regarding confined vs. eruptive events are limited by a small number of confined events that we analyzed (7 events). Extension of this work to a larger sample of confined and eruptive events would be necessary to draw more confident conclusions.
 








\section{Conclusions}\label{conc}
In this paper we analyzed the pre-flare vector magnetic fields within {\it AR, flare ribbons and PIL} areas in $40$ events and ARMS eruption simulation with an objective of improving our understanding of the physical properties of flaring vs. non-flaring photospheric vector magnetic fields in the photosphere.

Our quantitative findings are as follows:
\begin{itemize}
\item{{\it Ribbon vs.  PIL  morphology:} Qualitatively, comparing the near-PIL areas and flare ribbons in observations and simulations (Figures~\ref{bz_examples} and~\ref{fig_arms2}), we find that the morphology of PILs has almost no bearing on the morphology of flare ribbons. Hence, while one can accept \citet{Schrijver2007} contention that larger amounts of near-PIL flux indicate a greater likelihood of a larger flare, the spatial arrangement of the PIL does not substantively constrain ribbons’ spatial distribution.}


\item{ {\it Magnetic fluxes:} We find strong statistical correlations between the flare peak X-ray flux and the flare ribbon and PIL fluxes (Figure~\ref{4by4}a): Spearman correlation coefficient $r_s(I_\mathrm{X,peak}, \Phi_\mathrm{ribbon}) = 0.9$ and $r_s(I_\mathrm{X,peak}, \Phi_\mathrm{PIL}) = 0.6$, respectively in agreement with \citet{Kazachenko2017} analysis.  While correlation between flare size and PIL flux is in line with \citet{Schrijver2007} hypothesis that flare-productive ARs tend to have stronger PILs, it could reflect the fact that larger ARs have stronger PILs and hence host larger flares. The correlation between the peak X-ray flux and the corresponding AR quantities is weaker, ranging from $r_s(I_\mathrm{X,peak}, \Phi_\mathrm{AR}) = 0.3$ for full dataset ($n=3137$, \verb+RibbonDB+, \citealt{Kazachenko2017}) to $r_s(I_\mathrm{X,peak}, \Phi_\mathrm{AR}) = 0.6$ for \verb+FlareMagDB+ ($n=40$).

\item{\it Reconnection flux fractions:} We find moderate correlation between the flare peak X-ray flux and the fraction of AR magnetic flux participating in the flare, consistent with earlier results (Figure~\ref{4by4}b).
\item {\it Mean magnetic shears:} In both observations and simulations we find that the mean magnetic shear is strongest within PIL areas, decreasing gradually within ribbon and AR areas (Figure~\ref{4by4}c):
$\alpha(AR,rbn,PIL)=[10,18,26]\times10^3 G\cdot deg$ for \verb+RibbonDB+ vs. $\alpha(AR,rbn,PIL)=[0.06,0.13,0.17]\times10^3 G\cdot deg$ for ARMS.
{The peak X-ray flux is moderately correlated with the mean magnetic shear within the AR and is weakly correlated with the mean magnetic shear within ribbon and PIL areas. }

\item{\it Current density morphology:} Qualitatively, current density maps consist of thread-like and patchy structures that do not exhibit any regular shape and do not correlate with ribbon locations, flare and/or CME occurrence (Figure~\ref{jz_examples}).

\item {\it Total vertical currents:} We find that the total area-integrated unsigned vertical current is largest within AR areas decreasing gradually within ribbon and PIL areas (Figure~\ref{4by4}d). 

\item { {\it Net currents:} {\it Over entire active region}, we find that  currents are neutralized, in agreement with earlier studies (orange symbols in Figure~\ref{2by2}a): 
\begin{equation}
(DC/RC)_\mathrm{AR}=0.
\end{equation}
{\it Over one polarity (positive or negative)} currents are non-neutralized (blue symbols in Figure~\ref{2by2}a):
\begin{equation}
(DC/RC)_\mathrm{AR,+}\approx -(DC/RC)_\mathrm{AR,-}\ne0.
\end{equation}
The central part of the active region around the PIL has the highest net current decreasing gradually within ribbon and AR areas (see red, green and blue symbols, Figure~\ref{2by2}a). 
We find that the net current within flare ribbons strongly correlates with the mean magnetic shear within the PIL (Figure~\ref{2by2}b, $r_s=0.7$) with the scaling relationship $|DC/RC|\propto0.02\bar{\alpha}_\mathrm{PIL}+0.9$,  implying that current non-neutralization is a manifestation of the shear accumulation along the PIL in agreement with simulations \citep{Torok2014,Dalmasse2015}. }

\item{\it Confined vs. eruptive flares:}  We find that for a given peak X-ray flux, confined events have larger PIL fluxes and lower mean PIL shears and ribbon net currents  than eruptive events (see $\bigtriangleup$ vs. $\bullet$ in  Figures~\ref{4by4}a, c and d), in agreement with \citealt {Liu2017,Avallone2020}.  We also find that the CME speed has a strong correlation with the fraction of the AR that participates in the flare (Figure~\ref{3by1}). In fact the flux ratio is the only variable that has a correlation coefficient with the CME speed above $r_s>0.4$.}  

\end{itemize}
 
To summarize, following \citealt{Welsch2009} ``intensive-extensive'' classification, our analysis suggests that while flare peak X-ray fluxes are guided by {\it extensive} magnetic field properties that scale with the AR size (like the total amount of magnetic flux that participates in the reconnection process, \citealt{Bobra2015}), the CME speeds are guided by {\it intensive} properties that do not scale with the AR size (like the fraction between the reconnection flux and the AR flux, defined by the amount of overlying field, \citealt{Bobra2016,Sun2015}) with little dependence on the amount of mean PIL shear or net current.

This study is the largest-yet statistical analysis of the flare vector magnetic fields within flare-ribbon and PIL areas and the relationship with other flare and AR properties. Such a statistical approach is useful since it allows us to discover general laws that may be overlooked by individual case studies.
 
The \verb+FlareMagDB+ catalog is available online\footnote{http://solarmuri.ssl.berkeley.edu/~kazachenko/FlareMagDB/} in CSV and IDL sav file formats, along with maps of vector magnetic fields, magnetic shear, current densities, PIL and ribbon masks, and can be used for different types of quantitative studies in the future, from constraining the properties of the simulations to further detailed observation studies. For example, a comparison of magnetic field properties with presence of sigmoidal structures or filaments would be valuable to clarify the relationship between the coronal structures and the photospheric magnetic fields. Analysis of the outliers in the derived trends, for example, events with large magnetic shear and small net current and vice versa, would be interesting. Extension of this statistical work to a larger number of confined flares observed over a decade of SDO, including temporal evolution of the magnetic field properties over the flare (e.g. \citealt{Sharykin2020,Barczynski2020}), would allow us to further advance our understanding of solar eruption magnetism.
 
\begin{acknowledgments}
We thank the HMI team for providing us with the vector magnetic field SDO/HMI data. We thank Marc DeRosa and the AIA team for providing us with the SDO/AIA data.  We thank US taxpayers for providing the funding that made this research possible. We acknowledge support from NASA LWS NNH17ZDA001N, 80NSSC19K0070, NASA 80NSSC18K1283-HSR, NASA ECIP
NNH18ZDA001N (MDK), NASA LWS Award 80NSSC19K0072 (BTW), NASA award SV0-09020 (XS), NSF award 1848250 (XS), NASA HSR Award 80NSSC18K1283 (AS), and NASA Grand Challenges 996790 and 997022 (AS).
\end{acknowledgments}

{\bf DATA:  Dataset is available online at \verb+http://solarmuri.ssl.berkeley.edu/~kazachenko/FlareMagDB/+}


\appendix\label{appendix}

\section{FlareMagDB for all events: magnetic shear and vertical current maps}\label{app01}


\begin{figure*}[htb!]
  \centering  
\resizebox{1.0\hsize}{!}{\includegraphics[angle=0,width=\textwidth]{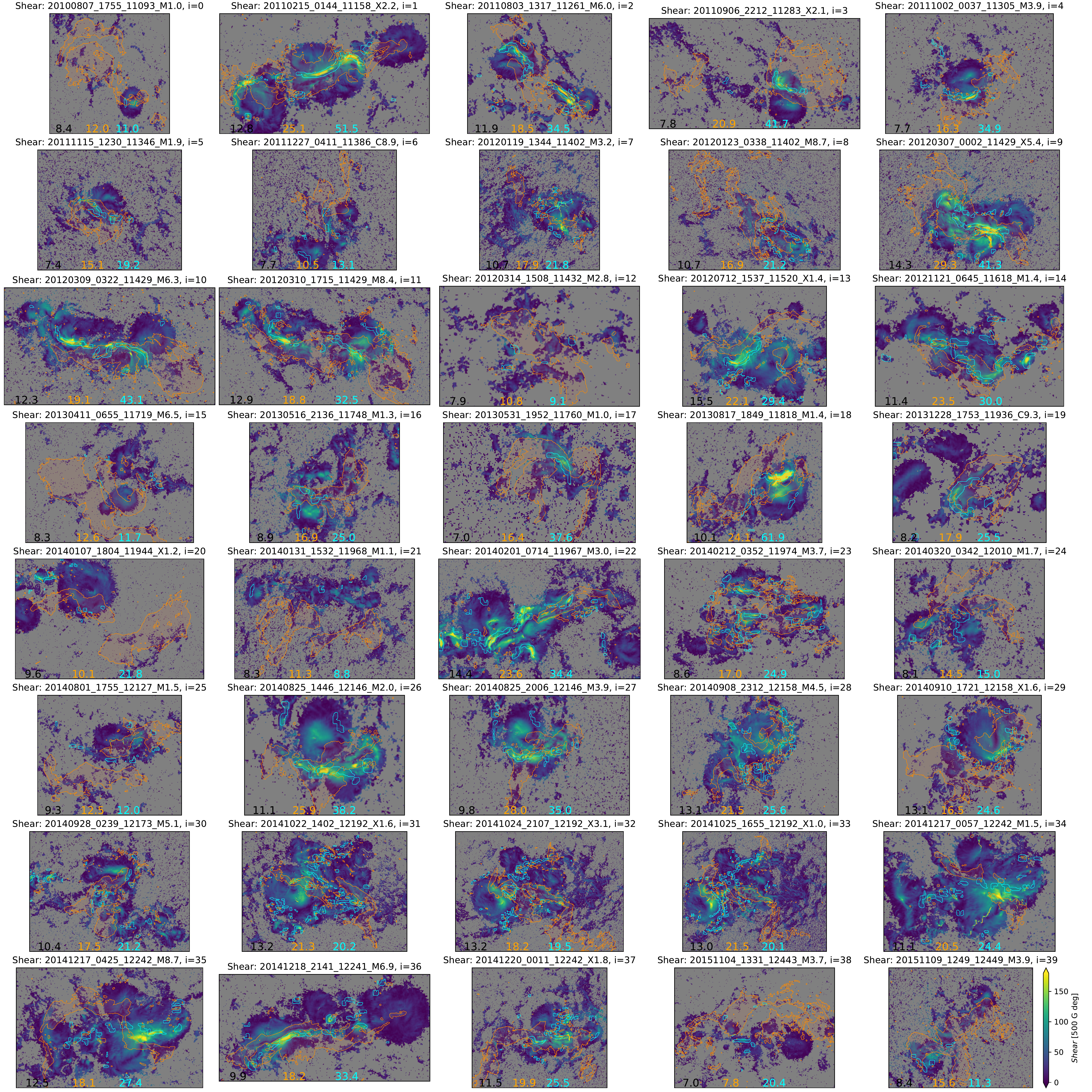}}  \caption{Magnetic field shear, $\alpha$, for $40$ events from the FlareMagDB database. Black, orange and cyan numbers indicate mean magnetic shear within AR, ribbon and PIL areas, respectively (see column $\bar\alpha$ in Table~\ref{events}).}
  \label{shear_examples}
\end{figure*}

\begin{figure*}[htb!]
  \centering  
\resizebox{1.0\hsize}{!}{\includegraphics[angle=0,width=\textwidth]{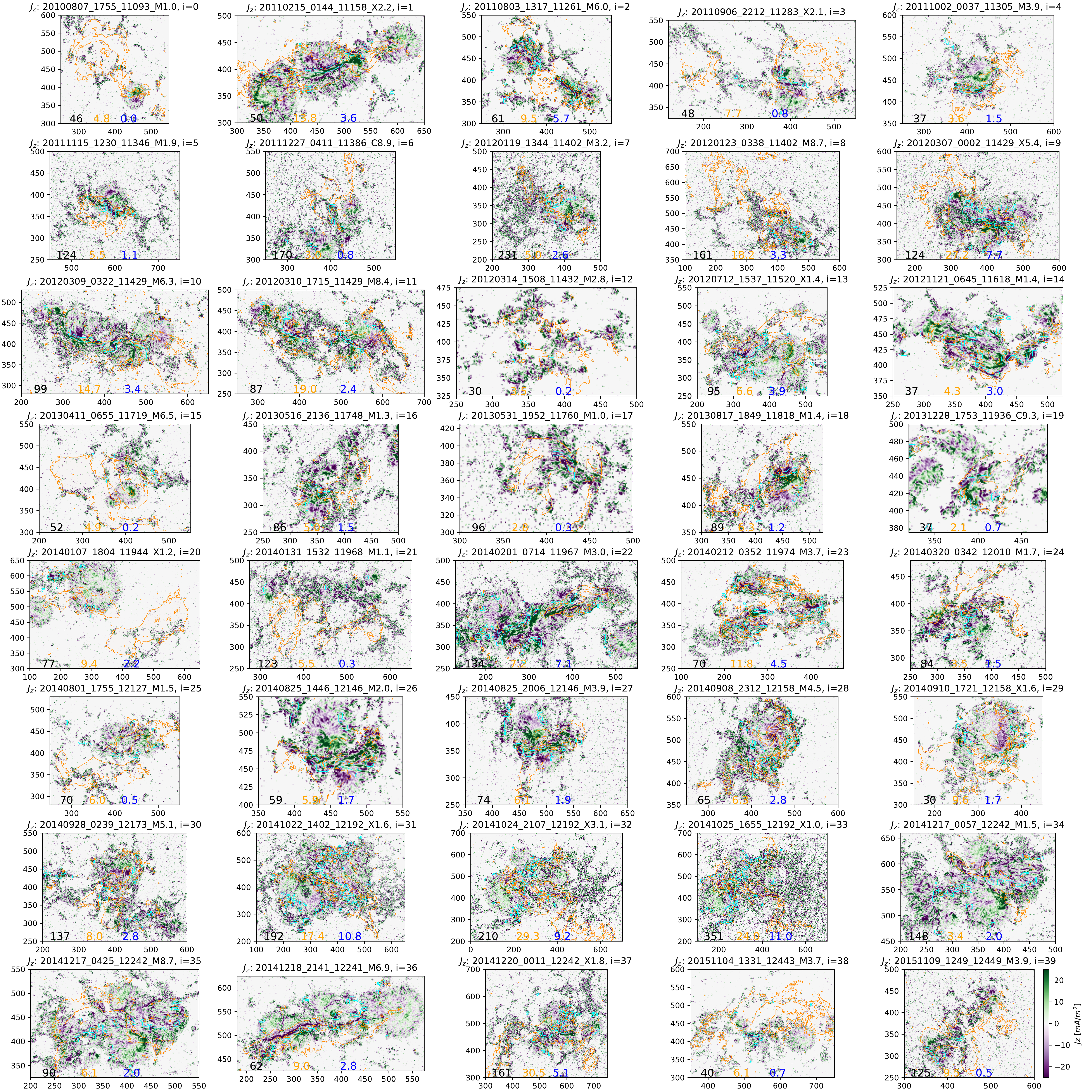}}  \caption{Vertical current distribution, $J_z$,  for $40$ events from the FlareMagDB database. Black, orange and blue numbers indicate total electric current within AR, ribbon and PIL areas, respectively (see column $I_\mathrm{z,u}$ in Table~\ref{events}).}
  \label{jz_examples}
\end{figure*}

\end{document}